\documentclass[nojss]{jss}

\usepackage{epsfig}
\usepackage{amsmath}
\usepackage{amssymb}

\newcommand{\real}{\mathbb{R}}
\newcommand{\I}{\mathbb{I}}
\newcommand{\rmd}{\mathrm{d}}
\newcommand{\N}{\mathrm{N}}
\newcommand{\cov}{\mathrm{cov}}

\author{Benjamin M. Taylor\\Lancaster University, UK \And 
      Tilman M. Davies\\Massey University, NZ \AND 
      Barry S. Rowlingson \\Lancaster University, UK \And 
      Peter J. Diggle \\Lancaster University, UK}
\title{\pkg{lgcp} -- An \proglang{R} Package for Inference with Spatio-Temporal
Log-Gaussian Cox Processes}

\Plainauthor{Benjamin M. Taylor, Tilman M. Davies, Barry S. Rowlingson, Peter J.
Diggle} %% comma-separated
\Plaintitle{lgcp -- An R Package for Inference with Spatio-Temporal Log-Gaussian
Cox Processes} %% without formatting
\Abstract{
This paper introduces an \proglang{R} package for spatio-temporal prediction and forecasting for log-Gaussian Cox processes. The main computational tool for
these models is Markov chain Monte Carlo and the new package, \pkg{lgcp},
therefore also
   provides an extensible suite of functions for implementing MCMC 
   algorithms for processes of this type. 
   The modelling framework and details of inferential procedures are first
presented before a tour of \pkg{lgcp} functionality is given via a walk-through
data-analysis. Topics covered include reading in and converting data, estimation
of the key components and parameters of the model, specifying output and
simulation quantities, computation of Monte Carlo expectations, 
   post-processing and simulation of data sets.
}
\Plainkeywords{spatio-temporal point processes, R} %% without formatting
\Address{
   \textbf{Benjamin M. Taylor}\\
   Division of Medicine,\\
   Lancaster University,\\
   Lancaster,\\
   LA1 4YF,\\
   UK\\
   E-mail: \email{b.taylor1@lancaster.ac.uk}\\
   URL: \url{http://www.maths.lancs.ac.uk/~taylorb1/}

   \textbf{Tilman M. Davies}\\
   Institute of Fundamental Sciences\\
   Massey University\\
   Science Tower ScB2.10\\
   Manawatu Campus\\
   Private Bag 11222\\
   Palmerston North 4442\\
   New Zealand\\
   E-mail: \email{T.Davies@massey.ac.nz}\\
   URL: \url{http://ifs.massey.ac.nz/people/staff.php?personID=232}

   \textbf{Barry Rowlingson}\\
   Division of Medicine,\\
   Lancaster University,\\
   Lancaster,\\
   LA1 4YF,\\
   UK\\
   E-mail: \email{b.rowlingson@lancaster.ac.uk}\\
   URL: \url{http://www.maths.lancs.ac.uk/~rowlings/}

   \textbf{Professor Peter J. Diggle}\\
   Division of Medicine,\\
   Lancaster University,\\
   Lancaster,\\
   LA1 4YF,\\
   UK\\
   E-mail: \email{p.diggle@lancaster.ac.uk}\\
   URL: \url{http://www.lancs.ac.uk/~diggle/}
}
\begin{document}

%% include your article here, just as usual
%% Note that you should use the \pkg{}, \proglang{} and \code{} commands.
%%
%%\section[About Java]{About \proglang{Java}}
%% Note: If there is markup in \(sub)section, then it has to be escape as above.

\section{Introduction}
\label{sect:intro}

This article introduces a new \proglang {R} package,
\pkg{lgcp}, 
for inference with spatio-temporal log-Gaussian Cox processes \cite{Rpackage}. The work was
motivated by applications in disease surveillance, where the major focus of
scientific interest is
on whether, and if so where and when, 
cases form unexplained clusters within a spatial region $W$
and time-interval $(0,T)$
of interest. 
It will be assumed that both the location and time of each case is known, at
least to a sufficiently fine resolution that a point process
modelling framework is natural. In general, the aims of
statistical analysis include model formulation,
parameter estimation and spatio-temporal
prediction. The \pkg{lgcp} package includes some functionality
for parameter estimation and diagnostic checking, mostly by linkages
with other \proglang {R} packages for spatial statistics. However, and
consistent with the scientific focus being on 
disease surveillance, the current version of the  package 
places particular emphasis on real-time predictive inference. Specifically,
using the modelling framework of a Cox process 
with stochastic intensity $R(x,t) = \exp\{{\mathcal Y}(x,t)\}$, where
${\mathcal Y}(x,t)$ is a Gaussian process, the package enables
the user to draw samples from the joint predictive distribution
of $R(x,t+k)$ given data up to and including time $t \leq T$,
for any set of locations $x \in W$ and forecast lead
time $k \geq 0$.
In the surveillance setting, these samples would typically be used to 
evaluate the predictive probability that the intensity at a
particular location and time exceeds a specified intervention
threshold; see, for example, \cite{diggle2005}.

The \pkg{lgcp} package makes extensive use of \pkg{spatstat} 
functions and data structures \citep{baddeley2005}.
Other  important dependencies are:
the \pkg{sp} package, which also supplies some data structures and functions
\citep{pebesma2005,bivand2008}; the suite of covariance functions 
provided by the \pkg{RandomFields} package \citep{schlather2001}; the
\pkg{rpanel} package to
facilitate  minimum-contrast parameter estimation 
routines \citep{bowman2007,bowman2010}; and 
the \pkg{ncdf} package for rapid access to massive 
datasets for post-processing  \citep{pierce2011}.

In Section \ref{sect:stlgcp} the log-Gaussian Cox process is introduced.
Section \ref{sect:inference} gives a review on methods
of inference for log-Gaussian Cox processes.
Section \ref{sect:packageintro} is an overview of the
package by way of a walk-through example,
covering: reading in data (Section \ref{sect:readdata}); 
estimating components of the model and associated parameters (Sections
\ref{sect:lambdaandmu} and \ref{sect:parest});
setting up and running the model (Sections \ref{sect:lgcppredict} and
\ref{sect:running}); and post-processing of command outputs (Section
\ref{sect:postprocessing}).
Some possible extensions of the package are given in Section
\ref{sect:extensions}. The appendices give
further information on the rotation of observation windows (Appendix
\ref{sect:rotation}), simulation of data (Appendix \ref{sect:simulation}) and
information about the \code{spatialAtRisk} class of objects (Appendix
\ref{sect:spatialAtRisk}), which may be useful for reference purposes.

\section[Spatio-Temporal log-Gaussian Cox processes]{Spatio-Temporal log-Gaussian Cox processes}
\label{sect:stlgcp}

Let $W\subset\real^2$ be an observation window in space and
$T\subset\real_{\geq 0}$ be an interval of time of interest. Cases 
occur at spatio-temporal positions $(x,t) \in W \times T$ 
according to an inhomogeneous spatio-temporal Cox process,
i.e., a Poisson process with a stochastic intensity $R(x,t)$.
The number of cases, $X_{S,[t_1,t_2]}$, arising in 
any $S \subseteq W$ during the interval $[t_1,t_2]\subseteq T$ is 
then Poisson distributed conditional on $R$,
\begin{equation}\label{eqn:themodel}
   X_{S,[t_1,t_2]} \sim \text{Poisson}\left\{\int_S\int_{t_1}^{t_2} R(s,t)\rmd
s\rmd t\right\}.
\end{equation}
Following \cite{diggle2005}, the intensity is decomposed multiplicatively as,
\begin{equation}
   R(s,t) = \lambda(s)\mu(t)\exp\{\mathcal Y(s,t)\}.
\label{eq:multiplicative}
\end{equation}
In Equation~\ref{eq:multiplicative},the
\emph{fixed spatial component}, $\lambda:\real^2\mapsto\real_{\geq 0}$, is a
known function, proportional to the population at risk at each point in space
and scaled so that,
\begin{equation}\label{eqn:intlambdas}
   \int_W\lambda(s)\rmd s=1,
\end{equation}
whilst the
\emph{fixed temporal component}, 
$\mu:\real_{\geq 0}\mapsto\real_{\geq 0}$, is also a known function 
such that,
\begin{equation}\label{eqn:mutdef}
   \mu(t) = \lim_{\delta t \rightarrow 0}
   \left\{\frac{\E[X_{W,\delta t}]}{ |\delta t|}\right\}.
\end{equation}

The function $\mathcal Y$ is a Gaussian process, continuous in both space and
time. In the nomenclature of epidemiology, the components $\lambda$ and $\mu$
determine the \emph{endemic} spatial and temporal component of the population at
risk; whereas $\mathcal Y$ captures the residual variation, or the
\emph{epidemic} component. 

The Gaussian process, $\mathcal Y$, is second order stationary with
minimally-parametrised covariance function,
\begin{equation}
\cov[\mathcal Y(s_1,t_1),\mathcal Y(s_2,t_2)] =
\sigma^2r(||s_2-s_1||;\phi)\exp\{-\theta(t_2-t_1)\},
\label{eq:cov}
\end{equation}
where $||\,\cdot\,||$ is a suitable norm on $\real^2$, for instance the
Euclidean norm, and $\sigma,\phi,\theta>0$ are known parameters. In the
\pkg{lgcp} package, the isotropic spatial correlation function, $r$, may take
one of several forms and possibly require additional parameters (in \pkg{lgcp}, this can be selected from any of the compatible models in the function \code{CovarianceFct} from the \pkg{RandomFields} package). The parameter $\sigma$ scales the log-intensity, 
whilst the parameters $\phi$ and $\theta$  govern the rates 
at which the correlation function decreases in space and in time,
respectively. The mean of the process $\mathcal Y$ is set equal to $-\sigma^2/2$
so as to give $\E[\exp\{\mathcal Y\}]=1$, hence the endemic/epidemic analogy above.

\section{Inference}
\label{sect:inference}

As in \cite{moller1998}, \cite{brix2001} and \cite{diggle2005}, a discretised version of the above model will be considered, defined on a regular grid
over space and time. Observations, $X$, are then
treated as cell counts on this grid. 
The discrete version of $\mathcal Y$ will be denoted $Y$; since $Y$ is a
finite collection of random variables, the properties of $\mathcal Y$ imply that
$Y$ has a multivariate Gaussian density with 
approximate covariance matrix $\Sigma$,
whose elements are calculated by
evaluating
Equation~\ref{eq:cov} at the centroids of the spatio-temporal
grid cells. Without loss of generality, unit time-increments are assumed
and events can be thought of as occurring ``at'' integer times $t$.
Let $X_t$ denote an observation over the spatial grid at time $t$,
and  $X_{t_1:t_2}$ denote the observations at times $t_1,t_1+1,\ldots,t_2$. 
For predictive
inference about $Y$,
samples from the conditional distribution
of the latent field, $Y_t$, given the 
observations to date, $X_{1:t}$ would be drawn,  but this is infeasible
because the dimensionality of the required 
integration increases without limit as time progresses.
An alternative, as suggested by \cite{brix2001}, is to sample from $Y_{t_1:t_2}$ given $X_{t_1:t_2}$,
\begin{equation}\label{eqn:jointdens}
   \pi(Y_{t_1:t_2}|X_{t_1:t_2}) \propto
\pi(X_{t_1:t_2}|Y_{t_1:t_2})\pi(Y_{t_1:t_2}),
\end{equation}
   where
$t_1 = t_2 - p$ for some small positive integer $p$. 
The justification for this approach is 
that observations from the remote past have a negligible
effect on inference for the current state, $Y_t$. 

In order to evaluate $\pi(Y_{t_1:t_2})$ in Equation~\ref{eqn:jointdens},
the parameters of the process $Y$ must either be known or estimated from the
data. Estimation of $\sigma$, $\phi$ and $\theta$ may be achieved either in a
Bayesian framework, or by 
one of a number of more {\it ad hoc} methods. 
The methods implemented in 
the current version of the \pkg{lgcp} package 
fall into the latter category and 
are described in \cite{brix2001} and \cite{diggle2005}. Briefly, this
involves matching empirical and theoretical estimates of the 
second-moment properties 
of the model. For the
spatial covariance parameters
$\sigma$ and $\phi$,
the inhomogeneous $K$-function, or $g$ function are used \citep{baddeley2000}. 
The autocorrelation function of the total 
event-counts per unit time-interval is used for  estimating the 
temporal correlation parameter $\theta$. The estimated parameter values can then be used 
to implement plug-in-prediction for the latent field $Y_t$.

\subsection[Discretising and the fast-Fourier transform]{Discretising and the
fast-Fourier transform}
\label{sect:fft}

The first barrier to inference is computation of the covariance matrix,
$\Sigma$, which even for relatively coarse grids is very large. Fortunately, for
regular spatial grids of size $2^m\times2^n$, there exist fast methods for
computing this based on the discrete Fourier transform \citep{wood1994}. The
general idea is to embed $\Sigma$ in a symmetric circulant matrix, $C=Q\Lambda
Q^*$, where $\Lambda$ is a diagonal matrix of eigenvalues of $C$, $Q$ is a
unitary matrix and
$^*$ denotes the Hermitian transpose. The entries of $Q$ are given by the
discrete Fourier transform. Computation of $C^{1/2}$, which is useful for both
simulation and evaluation of the density of $Y$, is then straightforward 
using the fact that $ C^{1/2} = Q\Lambda^{1/2} Q^*$.

\subsection[The Metropolis-adjusted Langevin algorithm]{The Metropolis-adjusted
Langevin algorithm}
\label{sect:mala}

Monte Carlo simulation from $\pi(Y_{t_1:t_2}|X_{t_1:t_2})$
is made more efficient
by working with a linear transformation of $Y$, partially determined by the
matrix $C$ as described below.
The  \pkg{lgcp} package
returns results
pertaining to $Y$
on a grid of size $M\times N\equiv 2^m\times2^n$ for positive integers 
$m$ and $n$ specified by the user,
which is extended to a grid of size $2M\times 2N$ for computation \citep{moller1998}.
Writing $\Gamma_t=\Lambda^{-1/2}Q(Y_t-\mu)$, the target of interest is given by,
\begin{equation}\label{eqn:target}
   \pi(\Gamma_{t_1:t_2}|X_{t_1:t_2}) \propto
\left[\prod_{t=t_1}^{t_2}\pi(X_t|Y_t)\right]\left[\pi(\Gamma_{t_1})\prod_{
t=t_1+1}^{t_2}\pi(\Gamma_{t}|\Gamma_{t-1})\right],
\end{equation}
where the first term on the right hand side of Equation~\ref{eqn:target} corresponds to
the first bracketed term on the right hand side of Equation~\ref{eqn:jointdens} and the
second bracketed term is the joint density, $\log\{\pi(\Gamma_{t_1:t_2})\}$.
Because
$\Gamma$ is
an Ornstein-Uhlenbeck process in time, the transition density,
$\pi(\Gamma_{t}|\Gamma_{t-1})$,
has an explicit expression as a Gaussian density; see \cite{brix2001}.

Since the gradient of the transition density
can also be written down explicitly, 
a natural and efficient MCMC method for sampling from the
predictive 
density of interest (Equation~\ref{eqn:target}),
is a Metropolis-Hastings algorithm with a Langevin-type proposal \citep{roberts1996},
\[
   q(\Gamma,\Gamma') = \N\left[\Gamma';\Gamma +
\frac12\nabla\log\{\pi(\Gamma|X)\},h^2\I\right],
\]
where $\N(y;m,v)$ denotes a Gaussian density with mean $m$ and variance $v$
evaluated at $y$,
$\I$ is the identity matrix and $h>0$ is a scaling parameter
\citep{metropolis1953,hastings1970}. 

Various theoretical results exist concerning the optimal acceptance probability
of the MALA (Metropolis-Adjusted Langevin Algorithm) -- see \cite{roberts1998}
and \cite{roberts2001} for example. In practical applications, the target
acceptance probability is often set to 0.574, which would be approximately
optimal for a Gaussian target as the dimension of the problem tends to infinity.
An algorithm for the automatic choice of $h$, so that this acceptance
probability is achieved without disturbing the ergodic property of the chain, is
also straightforward to implement, see \cite{andrieu2008}.

\section[Introducing the lgcp package]{Introducing the \pkg{lgcp} package}
\label{sect:packageintro}

\subsection{Reading-in and converting data}
\label{sect:readdata}

The generic data-format of interest is  $(x_i,y_i,t_i): i=1,...,n$, 
where the $(x_i,y_i)$ are the locations and 
$t_i$ the times of occurrence of  events  in $W \times (A,B)$, where
W is a polygonal observation window and $(A,B)$ the
time-interval within which events are observed. In the following
example \code{x}, \code{y} and \code{t} are \proglang{R} objects giving the
location and time of events and \code{win} is a \pkg{spatstat} object of class
\code{owin} specifying the polygonal observation window \citep{baddeley2005}. An example of constructing an appropriate \code{owin} object from ESRI shapefiles is given in the package vignette (type \code{vignette("lgcp")}).

\begin{CodeChunk}
\begin{CodeInput}
R> data <- cbind(x,y,t)
R> tlim <- c(0,100)
R> win
\end{CodeInput}
\begin{CodeOutput}
window: polygonal boundary
enclosing rectangle: [381.7342, 509.7342] x [64.14505, 192.14505] units
\end{CodeOutput}
\end{CodeChunk}

The first task for the user is to convert this into a space-time planar point
pattern object ie.\! one of class \code{stppp}, provided by \pkg{lgcp}. An object
of class \code{stppp} is easily created:
\begin{CodeChunk}
\begin{CodeInput}
R> xyt <- stppp(list(data=data,tlim=tlim,window=win))
R> xyt
\end{CodeInput}
\begin{CodeOutput}
Space-time point pattern
 planar point pattern: 10069 points 
window: polygonal boundary
enclosing rectangle: [381.7342, 509.7342] x [64.14505, 192.14505] units  
   Time Window : [ 0 , 100 ]
\end{CodeOutput}
\end{CodeChunk}

\subsection{Estimating the spatial and temporal component}
\label{sect:lambdaandmu}

There are many ways to estimate the fixed
spatial and temporal components of the log-Gaussian Cox process. The fixed
spatial component, $\lambda(s)$,
represents the spatial intensity of events,
averaged over time and scaled to integrate to 1 over the observation
window $W$. In epidemiological settings, this typically corresponds to
the spatial distribution of the population at risk, although
this information may not be directly available. 
The fixed temporal component, $\mu(t)$, is 
the mean number of events in $W$ per unit time.
Where the relevant demographic information is unavailable to
specify $\lambda(s)$ and $\mu(t)$ directly, 
\pkg{lgcp} provides basic functionality to estimate
them from the data.

\begin{figure}[htbp]
   \centering
   \includegraphics[width=0.5\textwidth,height=0.4\textwidth]{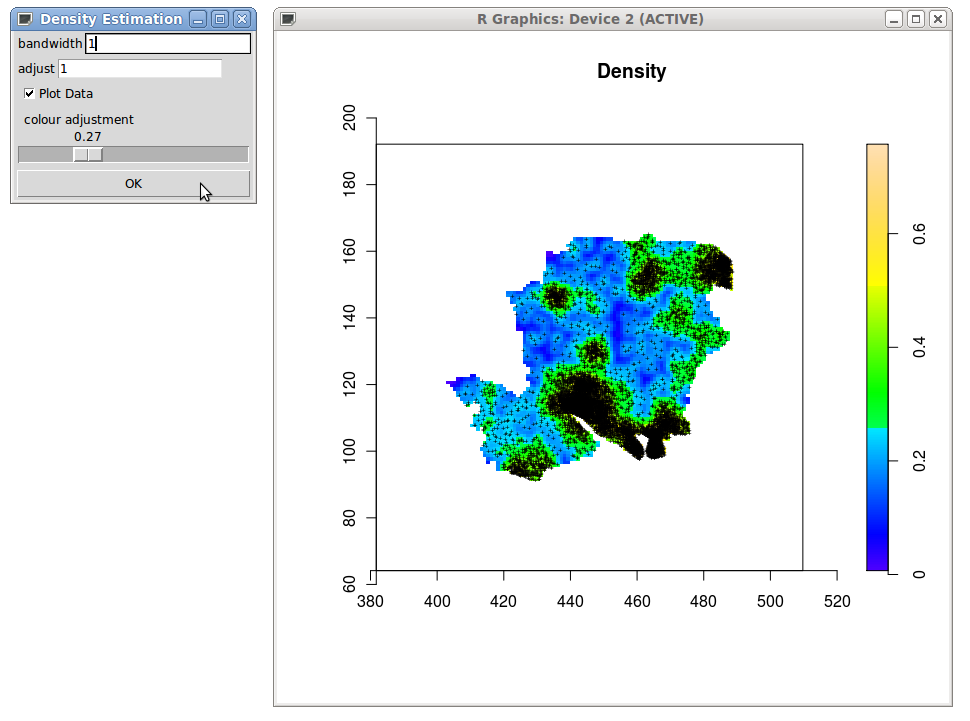}
   \caption{\label{lambdaEst} Choosing a kernel density estimate of
$\lambda(s)$.}
\end{figure}

The function \code{lambdaEst} is an interactive implementation
of a kernel method for estimating $\lambda(s)$
as in the following example:
\begin{CodeChunk}
\begin{CodeInput}
R> den <- lambdaEst(xyt,axes=TRUE)
R> plot(den)
\end{CodeInput}
\end{CodeChunk}
This calls an \pkg{rpanel} tool \citep{bowman2007} for estimating $\lambda$ (see Figure~\ref{lambdaEst}); once the user is happy
with the result, clicking on ``OK'' closes the panel and the kernel density estimate is stored
in the \proglang{R} object \code{den} of class \code{im} (a \pkg{spatstat} pixel image object). The estimate of
$\lambda$ can
then  be plotted in the usual way. The parameters \code{bandwidth} and \code{adjust} in this GUI relate to the arguments from the
\pkg{spatstat} function \code{density.ppp}; the former corresponds to the argument \code{sigma} and the latter to the argument of the same name.

The \pkg{lgcp} package
provides methods for coercing pixel-images like \code{den} to objects of class
\code{spatialAtRisk}, which can then
be used in parameter estimation and in the MALA algorithm to be discussed later; further useful information
on the \code{spatialAtRisk} class is provided in Appendix
\ref{sect:spatialAtRisk}.
\begin{CodeChunk}
\begin{CodeInput}
R> sar <- spatialAtRisk(den)
\end{CodeInput}
\begin{CodeOutput}
SpatialAtRisk object
   X range : [382.3742066569,509.094206656898]
   Y range : [64.785045372051,191.505045372051]
   dim     : 100 x 100
\end{CodeOutput}
\end{CodeChunk}
For the temporal component, $\mu(t)$, the user must provide an object that can
be coerced into one of class \code{temporalAtRisk}. 

\begin{figure}[htbp]
   \centering
   \includegraphics[width=0.4\textwidth,height=0.4\textwidth]{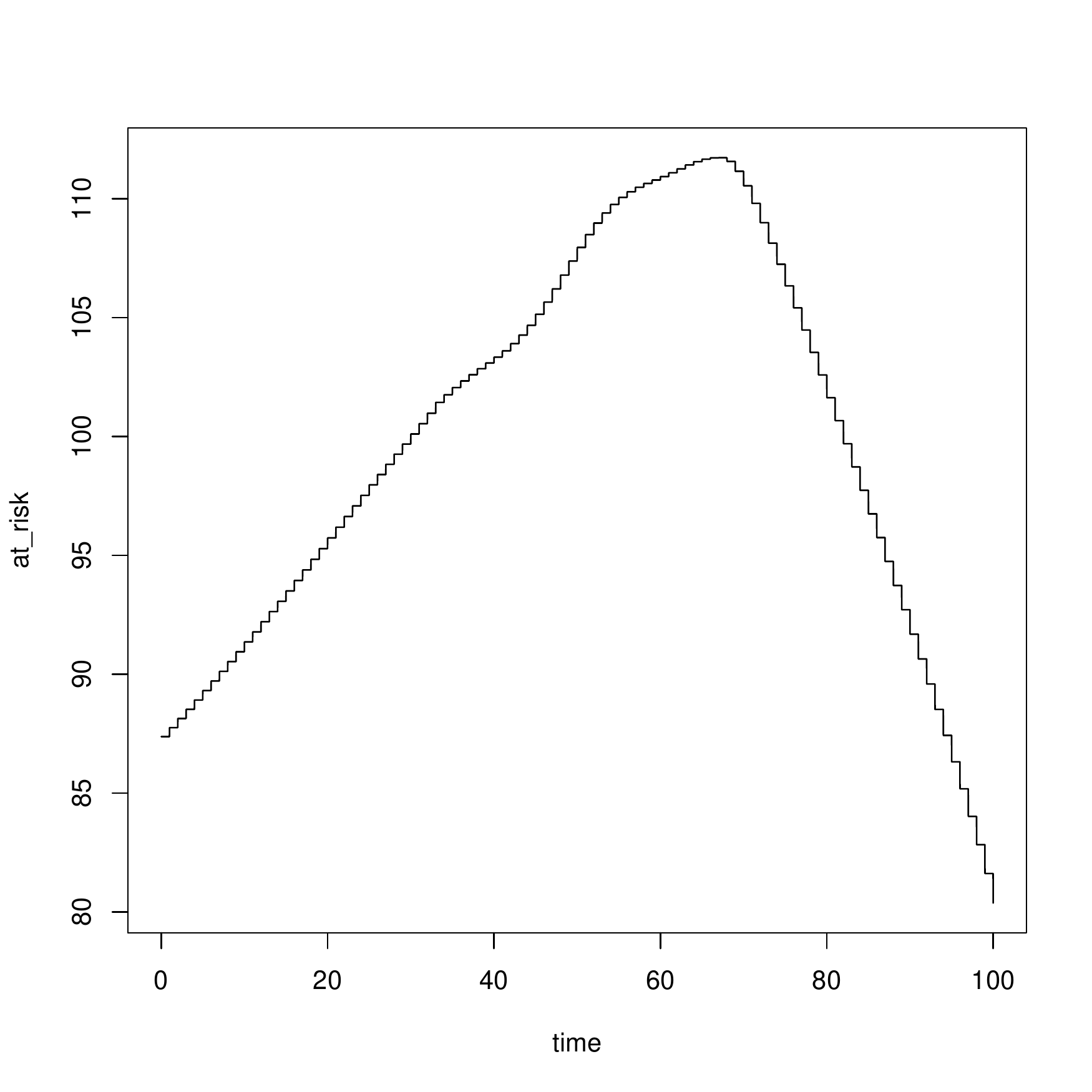}
   \caption{\label{muEst} Estimating $\mu(t)$: the output from
\code{plot(mut1)}.}
\end{figure}

Objects of class \code{temporalAtRisk} are non-negative functions of time over an observation time-window of interest, which must be the same as the time-window of the \code{stppp} data object, \code{xyt}. In some applications \citep{diggle2005}, $\mu(t)$  might represent the fitted values of a parametric model for the case counts over time. As it is not possible to provide generic functionality for parametric $\mu(t)$, a simple non-parametric estimate of  $\mu$ can be generated using the function
\code{muEst}:
\begin{CodeChunk}
\begin{CodeInput}
R> mut1 <- muEst(xyt)
R> mut1
\end{CodeInput}
\begin{CodeOutput}
temporalAtRisk object
   Time Window : [ 0 , 100 ]
\end{CodeOutput}
\end{CodeChunk}
In order to retain positivity, \code{muEst} fits a locally-weighted polynomial regression estimate (the \proglang{R} function \code{lowess}) to the square root of the interval counts and returns the square of this smoothed estimate (see Figure~\ref{muEst}). The amount of smoothing is controlled by the \code{lowess} argument \code{f} which specifies the proportion of points in the plot which influence the smoothed estimate at each value (see \code{?lowess}), for example \code{muEst(xyt,f=0.1)}. If 
the user wishes to specify a constant time-trend, $\mu(t)=\mu$,
the command
\begin{CodeChunk}
\begin{CodeInput}
R> mut <- constantInTime(xyt)
\end{CodeInput}
\end{CodeChunk}
returns the 
appropriate \code{temporalAtRisk} object, correctly scaled as in Equation~\ref{eqn:mutdef}.
The fixed temporal component can be also be supplied either as a vector or as a
function that is automatically coerced to a \code{temporalAtRisk} object and scaled
to achieve the condition in Equation~\ref{eqn:mutdef}.

\subsection{Estimating parameters}
\label{sect:parest}

After estimating $\lambda(s)$ and $\mu(t)$, the 
next step in the analysis is to estimate the 
covariance parameters of the process $\mathcal Y$. 
\pkg{lgcp} provides basic moment-based methods for this in the form of
\pkg{rpanel} GUIs that allow the user to choose $\sigma$,
$\phi$ and $\theta$ by eye \citep{bowman2007}. Parameter estimation by eye is both fast and reasonably robust and moreover emphasizes the fact that the underlying methods are \emph{ad hoc}. As mentioned above, it is possible to implement principled Bayesian parameter estimation for this model by integrating over the discretised latent-field, $Y$; this is a planned extension to the package (see Section \ref{sect:extensions}). 

\begin{figure}[htbp]
   \centering
   \begin{minipage}{0.5\textwidth}
      \includegraphics[width=\textwidth,height=0.8\textwidth]{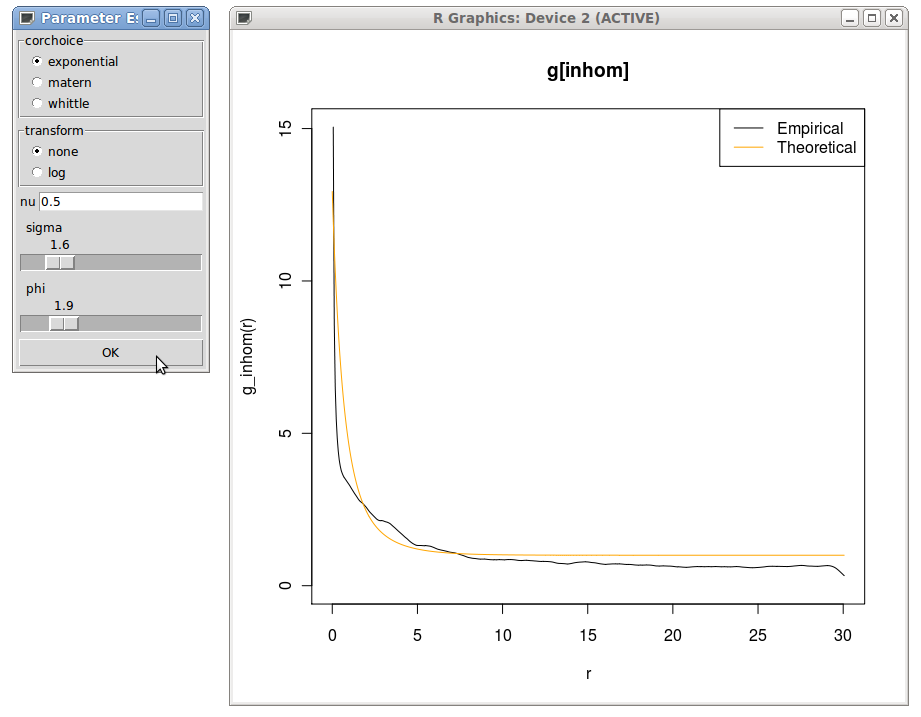}
   \end{minipage}\begin{minipage}{0.5\textwidth}
      \includegraphics[width=\textwidth,height=0.8\textwidth]{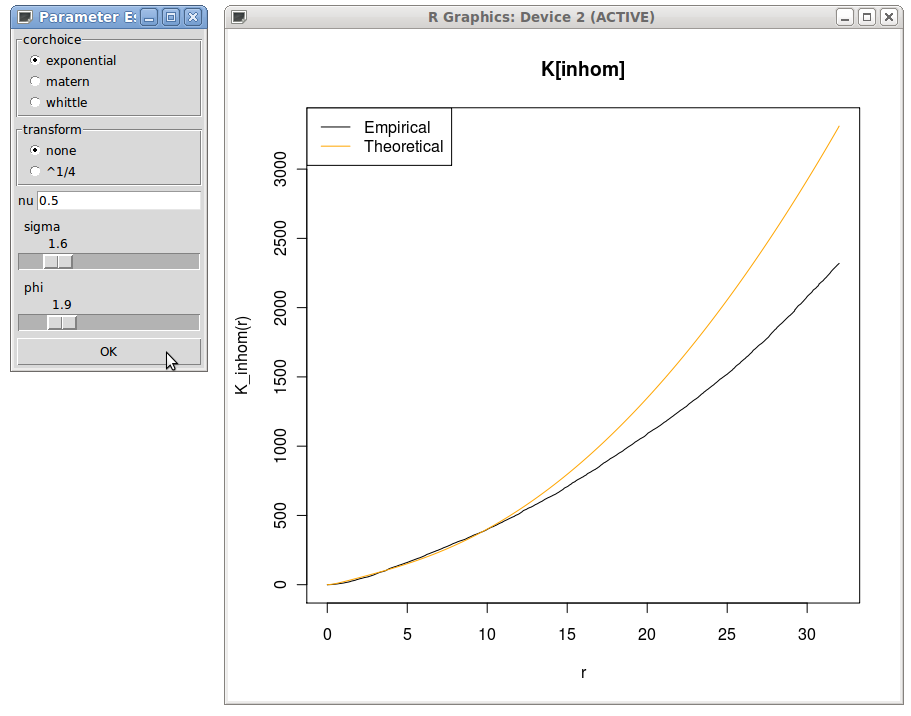}
   \end{minipage}
   \caption{\label{spatialparsEst} Estimating $\sigma$ and $\phi$: left via the
pair correlation function and right via the inhomogeneous $K$ function.}
\end{figure}

The spatial correlation parameters $\sigma$ and $\phi$ can be estimated
either from the pair correlation function, $g$, or the inhomogeneous $K$ function
\citep{baddeley2000}. 
Following \cite{brix2001} and \cite{diggle2005}, the
corresponding functions in \pkg{lgcp} estimate versions of
these two functions by averaging temporally localised
versions. The respective commands for doing so are:
\begin{CodeChunk}
\begin{CodeInput}
R> gin <- ginhomAverage(xyt,spatial.intensity=sar,temporal.intensity=mut)
R> kin <- KinhomAverage(xyt,spatial.intensity=sar,temporal.intensity=mut)
\end{CodeInput}
\end{CodeChunk}
The parameters are then estimated using either of the following:
\begin{CodeChunk}
\begin{CodeInput}
R> sigmaphi1 <- spatialparsEst(gin,sigma.range=c(0,10),phi.range=c(0,10))
R> sigmaphi2 <- spatialparsEst(kin,sigma.range=c(0,10),phi.range=c(0,10))
\end{CodeInput}
\end{CodeChunk}
These invoke another call to 
\pkg{rpanel}, which
produces the plots in Figure~\ref{spatialparsEst}.
The user's task is to match the orange 
theoretical function with the black
empirical counterpart.
The user can choose between the exponential, Mat\'ern or Whittle 
families of correlation functions; see \code{?CovarianceFct} from the
\pkg{RandomFields} package. For the 
Mat\'ern or Whittle 
families an additional shape parameter, $\nu$, is specified by the user 
and treated as a known constant. This is because
$\nu$ is known to be difficult to estimate. A recommended
strategy is to choose between a discrete set of candidate values;
for example, in the Mat\'ern family the integer part of $\nu$ gives
the number of times the underlying Gaussian process is mean-square
differentiable. The resulting estimated parameters are returned in list objects (e.g., \code{sigmaphi1} or \code{sigmaphi2}) with \code{sigmaphi1$sigma} and \code{sigmaphi1$phi} returning the required values of $\sigma$ and $\phi$. In the code below, note that these values have been input manually as respectively $1.6$ and $1.9$. The user has additional control over the  minimum contrast estimation, for example the range of evaluation, though sensible defaults are provided automatically by the embedded \pkg{spatstat} functions.

\begin{figure}[htbp]
   \centering
   \includegraphics[width=0.5\textwidth,height=0.4\textwidth]{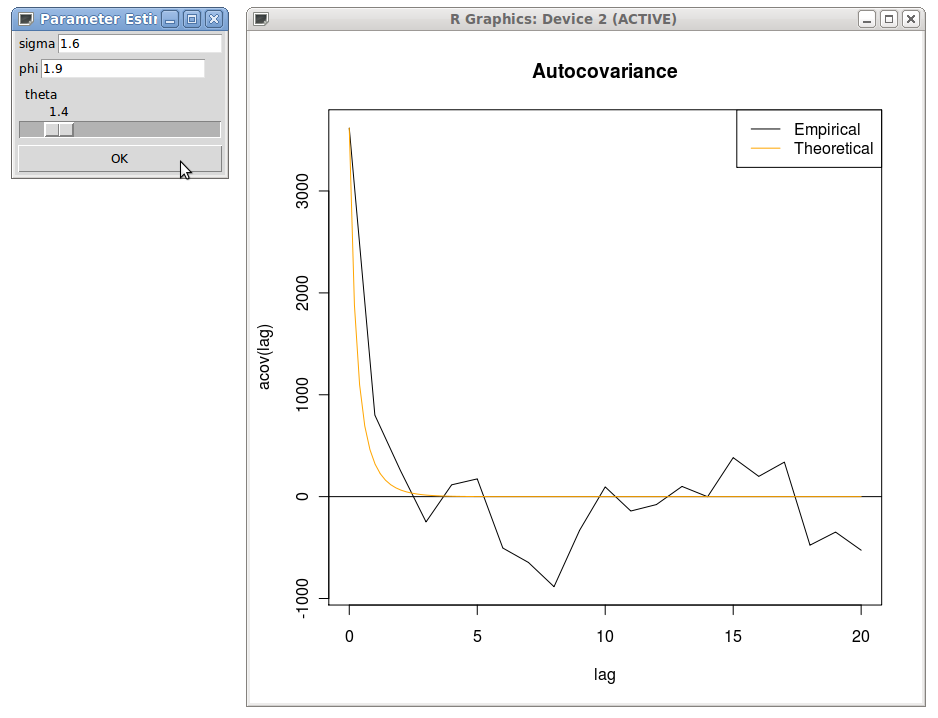}
   \caption{\label{thetaEst} Estimating $\theta$.}
\end{figure}

The temporal correlation parameter, $\theta$, 
can be
estimated using the function \code{thetaEst}; this requires $\sigma$, $\phi$ and $\mu(t)$
to have been estimated beforehand. For example, the call
\begin{CodeChunk}
\begin{CodeInput}
R> theta <- thetaEst(xyt,spatial.intensity=sar,
                        temporal.intensity=mut,sigma=1.6,phi=1.9)
\end{CodeInput}
\end{CodeChunk}
gives the result shown in Figure~\ref{thetaEst}. Note that again, in the code below the estimated value of $1.4$ has been input manually.

\subsection[The command lgcpPredict]{The command \code{lgcpPredict}}
\label{sect:lgcppredict}

The main function in the \pkg{lgcp} package is \code{lgcpPredict}.
This uses an
MCMC method to produce samples and
summary statistics from the predictive
distribution of the discretised
process $Y$, treating
previously obtained estimates
of $\lambda(s)$, $\mu(t)$, and the covariance
parameters as known quantities.
The MCMC algorithm
is invoked by the command \code{lgcpPredict}, 
whose arguments are as follows:
\begin{CodeChunk}
\begin{CodeInput}
R> args(lgcpPredict)
\end{CodeInput}
\begin{CodeOutput}
function (xyt, T, laglength, model.parameters = lgcppars(), 
   spatial.covmodel = "exponential", covpars = c(), cellwidth = NULL, 
   gridsize = NULL, spatial.intensity, temporal.intensity, mcmc.control, 
   output.control = setoutput(), autorotate = FALSE, gradtrunc = NULL) 
\end{CodeOutput}
\end{CodeChunk}

\subsubsection{Data and model specification}

The argument \code{xyt} is the \code{stppp} object that contains the data,
\code{T} is the time-point of interest for prediction
(c.f., the time $t_2$ in Section \ref{sect:inference})
and \code{laglength} tells the algorithm the number of previous time-points 
whose 
data should be included.

Model parameters are set using the
\code{model.parameters} 
argument; for example,
\begin{CodeChunk}
\begin{CodeInput}
R> lgcppars(sigma=1.6,phi=1.9,theta=1.4)
\end{CodeInput}
\end{CodeChunk}
has the obvious interpretation.
The spatial covariance model and
any additional parameters
are specified using the \code{spatial.covmodel} 
and \code{covpars} arguments; these may come from any of the compatible
covariance functions detailed in \code{?CovarianceFct} from the \pkg{RandomFields} package. The physical dimensions
of the grid cells can be set using either
the \code{cellwidth} or \code{gridsize} arguments,
the second of which
gives the number of cells in the $x$ and $y$ directions (these numbers are automatically extended to be a power of two for the fast-Fourier transform). 
The  \code{spatial.intensity} and \code{temporal.intensity} arguments
specify the previously obtained
estimates of $\lambda(s)$ and $\mu(t)$, respectively. 

It remains to set the MCMC parameters and output controls; these will now be
discussed.

\subsubsection{Controlling MALA and performing adaptive MCMC}

The \code{mcmc.control} argument of \code{lgcpPredict}
specifies the MCMC implementation and
is set using the \code{mcmcpars} function:
\begin{CodeChunk}
\begin{CodeInput}
R> args(mcmcpars)
\end{CodeInput}
\begin{CodeOutput}
function (mala.length, burnin, retain, inits = NULL, MCMCdiag = 0, 
   adaptivescheme)
\end{CodeOutput}
\end{CodeChunk}
Here, \code{mala.length} is the number of iterations to perform,
\code{burnin} is the number of iterations to throw away at the start
and
\code{retain} is the frequency at which to store or perform computations; 
for example, \code{retain=10} performs an action every 10th iteration. The
optional
argument \code{inits} can be used
to set initial values of $\Gamma$ for the algorithm,
and is intended for advanced use. The initial values are stored in a list object of length \code{laglength}$ + 1$, each element being a matrix of
dimension $2M\times2N$. The option \code{MCMCdiag} allows the user to save a sample of the $\Gamma$ chains; if \code{MCMCdiag=5} for example, then a
random sample of 5 MCMC chains from 5 different cells are stored; these can be
used to assess the convergence or mixing of the chain in the post-processing stage. 

The MALA proposal tuning parameter $h$ in Section \ref{sect:mala} must also be
chosen. The most straightforward way to do this is to set
\code{adaptivescheme=constanth(0.001)}, 
which gives $h=0.001$. Without a lengthy
tuning process, the value of $h$ that optimizes
the mixing of the algorithm
is not known. One solution to the problem of having to choose a scaling
parameter from pilot runs is to use adaptive MCMC \citep{roberts2009,andrieu2008}. Adaptive MCMC algorithms
use information from the realisation of an MCMC chain to make adjustments
to the proposal kernel. The Markov property is therefore
no longer satisfied and some care must be taken to ensure that the correct
ergodic distribution is preserved.
An elegant method, introduced by \cite{andrieu2008} uses a Robbins-Munro
stochastic approximation update to adapt the tuning parameter of the proposal
kernel \citep{robbins1951}. The idea is to update the tuning parameter at each
iteration of the sampler according to the iterative scheme,
\begin{equation}
   h^{(i+1)} = h^{(i)} + \eta^{(i+1)}(\alpha^{(i)} - \alpha_\text{opt}),
\end{equation}
where $h{(i)}$ and $\alpha^{(i)}$ are the tuning parameter and acceptance
probability at iteration $i$ and $\alpha_\text{opt}$ is
the target
acceptance probability. For Gaussian targets, and in the limit as the dimension
of the problem tends to infinity, an appropriate target acceptance probability
for MALA algorithms is 
$\alpha_\text{opt}=0.574$ \citep{roberts2001}. 
The sequence $\{\eta^{(i)}\}$ is chosen so that $\sum_{i=0}^\infty\eta^{(i)}$ is
infinite 
but for any $\epsilon>0$,
$\sum_{i=0}^\infty\left(\eta^{(i)}\right)^{1+\epsilon}$ is finite. 
These two conditions ensure that any value of $h$ can be reached, but in a way
that maintains the ergodic behaviour of the chain. One class of sequences with
this property is,
\begin{equation}
   \eta^{(i)} = \frac{C}{i^\alpha},
\end{equation}
where $\alpha\in(0,1]$ and $C>0$ \citep{andrieu2008}.

The tuning constants for this algorithm are set with the function
\code{andrieuthomsh}.
\begin{CodeChunk}
\begin{CodeInput}
R> args(andrieuthomsh)
\end{CodeInput}
\begin{CodeOutput}
function (inith, alpha, C, targetacceptance = 0.574)
\end{CodeOutput}
\end{CodeChunk}
In the above, \code{inith} is the initial value of $h$
and the remaining arguments correspond to their counterparts in 
the text above. 

The advanced user can also write their own adaptive scheme, detailed examples of which are provided in the package vignette. Briefly, writing an adaptive MCMC scheme involves writing two functions to tell \proglang{R} how to initialise and update the values of $h$. This may sound simple, but it is crucial that these functions preserve the correct ergodic distribution of the MCMC chain, an appreciation of these subtleties is \textbf{essential} before any attempt is made to code such schemes.

\subsubsection{Specifying output}
\label{sect:outputspec}

By default,
\code{lgcpPredict} computes the mean and variance of $Y$ and the mean and variance of $\exp\{Y\}$ (the relative risk) for each of the grid
cells and time intervals of interest. 
Additional storage and online computations are specified by the
\code{output.control} argument and the \code{setoutput} function:
\begin{CodeChunk}
\begin{CodeInput}
R> args(setoutput)
\end{CodeInput}
\begin{CodeOutput}
function (gridfunction = NULL, gridmeans = NULL) 
\end{CodeOutput}
\end{CodeChunk}
The option \code{gridfunction} is used to declare general operations perform during simulation (for example, dumping the simulated $Y$s to disk), whilst user-defined Monte Carlo averages are computed using \code{gridmeans}. A
complete run of the MALA chain can be saved using the \code{dump2dir} function:
\begin{CodeChunk}
\begin{CodeInput}
R> args(dump2dir)
\end{CodeInput}
\begin{CodeOutput}
function (dirname, lastonly = TRUE, forceSave = FALSE) 
\end{CodeOutput}
\end{CodeChunk}
The user supplies a character string, \code{dirname}, 
giving the name of a directory in which the results are to be saved. 
The other arguments to \code{dump2dir} are, respectively, an option to
save only the last grid (i.e., the time \code{T} grid) and to bypass a safety
message that would otherwise be
displayed when \code{dump2dir} is invoked. The safety message 
warns the user of disk space requirements for saving.
For example, on a $128\times128$ output grid using 5 days of data, 
1000 simulations from the MALA will take up approximately 625 megabytes.

Another option is to compute Monte Carlo expectations,
\begin{eqnarray}
   \E_{\pi(Y_{t_1:t_2}|X_{t_1:t_2})}[g(Y_{t_1:t_2})] &=& \int_W
g(Y_{t_1:t_2})\pi(Y_{t_1:t_2}|X_{t_1:t_2})\rmd Y_{t_1:t_2},\\
   &\approx& \frac1n\sum_{i=1}^n g(Y_{t_1:t_2}^{(i)})
\end{eqnarray}
where $g$ is a function of interest, $Y_{t_1:t_2}^{(i)}$ is the $i$th retained
sample from the target and $n$ is the total number of retained iterations. For
example, to compute the mean of $Y_{t_1:t_2}$, set
$g(Y_{t_1:t_2}) = Y_{t_1:t_2}$.
The output from such a Monte Carlo average would 
then
be a set of $t_2-t_1$ grids, each cell of which 
is equal to the mean over all retained iterations of the algorithm. In the
context of setting up the \code{gridmeans} option to compute the Monte Carlo mean, the user would 
define a function \code{g} as
\begin{CodeChunk}
\begin{CodeInput}
R> gfun <- function(Y){
       return(Y)
   }
\end{CodeInput}
\end{CodeChunk}
and input this to the MALA run
using the function \code{MonteCarloAverage},
\begin{CodeChunk}
\begin{CodeInput}
R> args(MonteCarloAverage)
\end{CodeInput}
\begin{CodeOutput}
function (funlist, lastonly = TRUE) 
\end{CodeOutput}
\end{CodeChunk}

Here, \code{funlist}
is either a list or a character vector giving 
the names of the function(s) $g$. The specific syntax for the example above
would be a call of the form \code{MonteCarloAverage("gfun")}. The functions of interest (e.g., \code{gfun} above) are assumed to act on each of the individual grids, $Y_{t_i}$, and return a grid of the same dimension.

A second example arises in
epidemiological studies where
of clinical interest to know 
whether,
at any location
$s$, the ratio of current to expected
risk exceeded a pre-specified intervention threshold; 
see, for example,  \cite{diggle2005}, where real-time
predictions of relative risk are presented as
maps of exceedance probabilities,
${\rm P}\{\exp(Y_t)>k|{X_{1:t}}\}$ for a pre-specified
   threshold $k$. Any such exceedance probability can 
   be expressed as an expectation,   
\[
\Prob[\exp(Y_{t_1:t_2})>k]=\E_{\pi(Y_{t_1:t_2}|X_{t_1:t_2})}[\I(Y_{t_1:t_2}>k)]
= \frac1n\sum_{i=1}^n\I(Y_{t_1:t_2}^{(i)}>k),
\]
where $\I$ is the indicator function, and a Monte Carlo
approximation can
   therefore be computed on-line using \code{MonteCarloAverage}.

The corresponding function $g$ is 
\[
   g(Y_{t_1:t_2}) = \I(Y_{t_1:t_2}>k).
\]
Exceedance probabilities are made available directly within
\pkg{lgcp} by the function \code{exceedProbs}.

To make use of this facility, 
the user specifies the thresholds of interest,
for example 1.5, 2 and 3,
then creates a function to compute the required exceedances:
\begin{CodeChunk}
\begin{CodeInput}
R> exceed <- exceedProbs(c(1.5,2,3))
\end{CodeInput}
\end{CodeChunk}
The object \code{exceed} is now a function that returns the exceedance
probabilities as an array object of dimension $M\times N\times3$.
This function can be passed through to the \code{gridmeans} option,
together with the previously defined \code{gfun}, via
\code{gridmeans=MonteCarloAverage(c("gfun","exceed")}. The
\code{lgcpPredict} function then returns point-wise 
predictive means and three sets of exceedance probabilities.
Note that, the example function \code{gfun} is included for illustrative purposes only and is in fact redundant,
as \code{lgcpPredict} automatically returns the predictive mean (and variance) of $Y$.

\subsubsection{Rotation}

Testing whether estimation can proceed more efficiently in 
a rotated space is described
in detail in Appendix \ref{sect:rotation}. Note that if the data and
observation window are rotated, then $\lambda$ must also be rotated
to retain compatibility. If $\lambda$ 
was estimated in the original frame of reference 
and \code{autorotate=TRUE}, then \code{lgcpPredict} 
will automatically rotate $\lambda$ if 
it is computationally worthwhile to do so. 
For $\lambda$ specified as a grid,
either directly or via an object of class \code{im}, 
then a small amount of information loss occurs in the rotation
because the square cells in the original orientation become misaligned with the axes in the rotated space and vice-versa. If $\lambda$ is specified
by a continuous function, then no such loss occurs.

\subsubsection{Gradient truncation}

One undesirable
property of the Metropolis-adjusted Langevin algorithm is that the chain is
prone to taking very long excursions from the mode of the target; this behaviour
can have a detrimental effect on the mixing of the chain and consequently on any
results. The tendency to make long excursions is caused by instability in the
computation of the gradient vector, but the issue is relatively straightforward
to rectify without affecting convergence properties \citep{moller1998}. The key
is to truncate the gradient vector if it becomes too large. If
\code{gradtrunc=NULL}, then an appropriate truncation is automatically selected
by the code. 

As far as the authors are aware, there are no 
published guidelines for selecting this truncation parameter.
The current version of the 
\code{lgcp}
package uses the maximum achieved gradient over a set of 100 independent
realisations of $\Gamma_{t_1:t_2}$.

\subsection{Running}
\label{sect:running}

When all of the above options have been specified, the MALA 
algorithm can be called as follows:
\begin{CodeChunk}
\begin{CodeInput}
R> lg <- lgcpPredict(xyt=xyt,
		     T=50,
		     laglength=5,
		     model.parameters=lgcppars(sigma=1.6,phi=1.9,theta=1.4),
		     cellwidth=2,
		     spatial.intensity=sar,
		     temporal.intensity=mut,			
		     mcmc.control=mcmcpars(mala.length=120000,burnin=20000,
		        retain=100,MCMCdiag=5,
		        adaptivescheme=andrieuthomsh(inith=1,alpha=0.5,C=1,
			   targetacceptance=0.574)),
		     output.control=setoutput(gridfunction=
			   dump2dir(dirname="C:/MyDirectory"),
		        gridmeans=MonteCarloAverage("exceed",lastonly=TRUE)))
\end{CodeInput}
\end{CodeChunk}

The above call assumes that the spatial covariance model 
is exponential, that no rotation is to be performed 
and that the user wishes to have \code{lgcpPredict} compute 
an appropriate gradient truncation automatically. The arguments
\code{spatial.intensity} and \code{temporal.intensity} relate to the spatial and temporal intensities, estimated in Section \ref{sect:lambdaandmu}; note that the chosen temporal model is constant in time. The option \code{lastonly=TRUE} in \code{MonteCarloAverage} has the effect of only performing the
computation on data from the last time point, 
to save disk space. A similar option is available
for the \code{dump2dir} function.

The simulated example uses data from times 45 to 50 inclusive, 
120,000 iterations, of which the first
20,000 are treated as burn-in,
and retains every 100th sample.
For diagnostic checking,
a sample of MCMC $\Gamma$-chains is also saved for each of five 
randomly selected grid-cells.
The observation window is approximately 100km square, 
so the specified cell width of 2km 
gives an output grid of size $64\times64$, 
i.e., computation is carried out on a $128\times128$ grid. 
The complete run is saved to disk and 
exceedance probabilities are computed for the last time-point only.

During simulation, a progress bar is displayed giving the percentage of
iterations completed.

\subsection{Post-processing}
\label{sect:postprocessing}

The stored output \code{lg} is an object of 
class \code{lgcpPredict}. Typing \code{lg} into the console prints out
information about the run:
\begin{CodeChunk}
\begin{CodeInput}
R> lg
\end{CodeInput}
\begin{CodeOutput}
lgcpPredict object.

General Information
-------------------
      FFT Gridsize: [ 128 , 128 ]

	    Data:
     Time |       45       46       47       48       49       50
   Counts |      103      346      108      100       76       69

      Parameters: sigma=1.6, phi=1.9, theta=1.4
   Dump Directory: C:/MyDirectory

   Grid Averages:
	 Function Output Class
	 exceed        array

      Time taken: 14.148 hours

MCMC Information
----------------
   Number Iterations: 120000
	    Burn-in: 20000
	    Thinning: 100
      Mean Acceptance: 0.567
      Adaptive Scheme: andrieuthomsh
	       Last h: 0.001
\end{CodeOutput}
\end{CodeChunk}
Information returned includes the FFT grid size used in computation; the count
data for each day; the parameters used; the directory,
if specified,
to which the simulation was dumped; 
a list of \code{MonteCarloAverage} functions together with the \proglang{R}
class of their returned values; the time taken to do the simulation; and
information on the MCMC run.

\subsubsection{Extracting information}

The cell-wise mean and variance of $Y$ computed via Monte Carlo can 
always be extracted using \code{meanfield(lg)} and \code{varfield(lg)},
respectively. The calls \code{rr(lg)}, \code{serr(lg)} and \code{intens(lg)} return respectively the Monte Carlo mean relative risk (the mean of $\exp\{Y\}$), the standard error of the relative risk and the estimated cell-wise mean Poisson intensity. The $x$ and $y$ coordinates for the grid output are obtained via
\code{xvals(lg)} and \code{yvals(lg)}. If invoked, the commands
\code{gridfun(lg)} and \code{gridav(lg)} return respectively the
\code{gridfunction} and \code{gridmeans} options of the \code{setoutput}
argument of the \code{lgcpPredict} function,
whilst \code{window(lg)} returns the observation window.

Note that the structure produced by \code{gav <- gridav(lg)} is a \code{list} of
length 2. The first element of \code{gav}, retrieved with \code{gav$names}, is a
list of the function names given in the call to \code{MonteCarloAverage}. The
second element, \code{gav$output}, is a list of the function outputs; the $i$th
element in the list being the output from the
function corresponding to the $i$th element of \code{gav$names}. To return the output for a specific function, use the syntax \code{gridav(lg,fun="exceed")}, which in this case returns the exceedance probabilities, for example.

\subsubsection{Plotting}

\begin{figure}[htbp]
   \centering
   \begin{minipage}{0.05\textwidth}
   \hfill
   \end{minipage}\begin{minipage}{0.5\textwidth}
      \includegraphics[width=0.8\textwidth,height=0.8\textwidth]{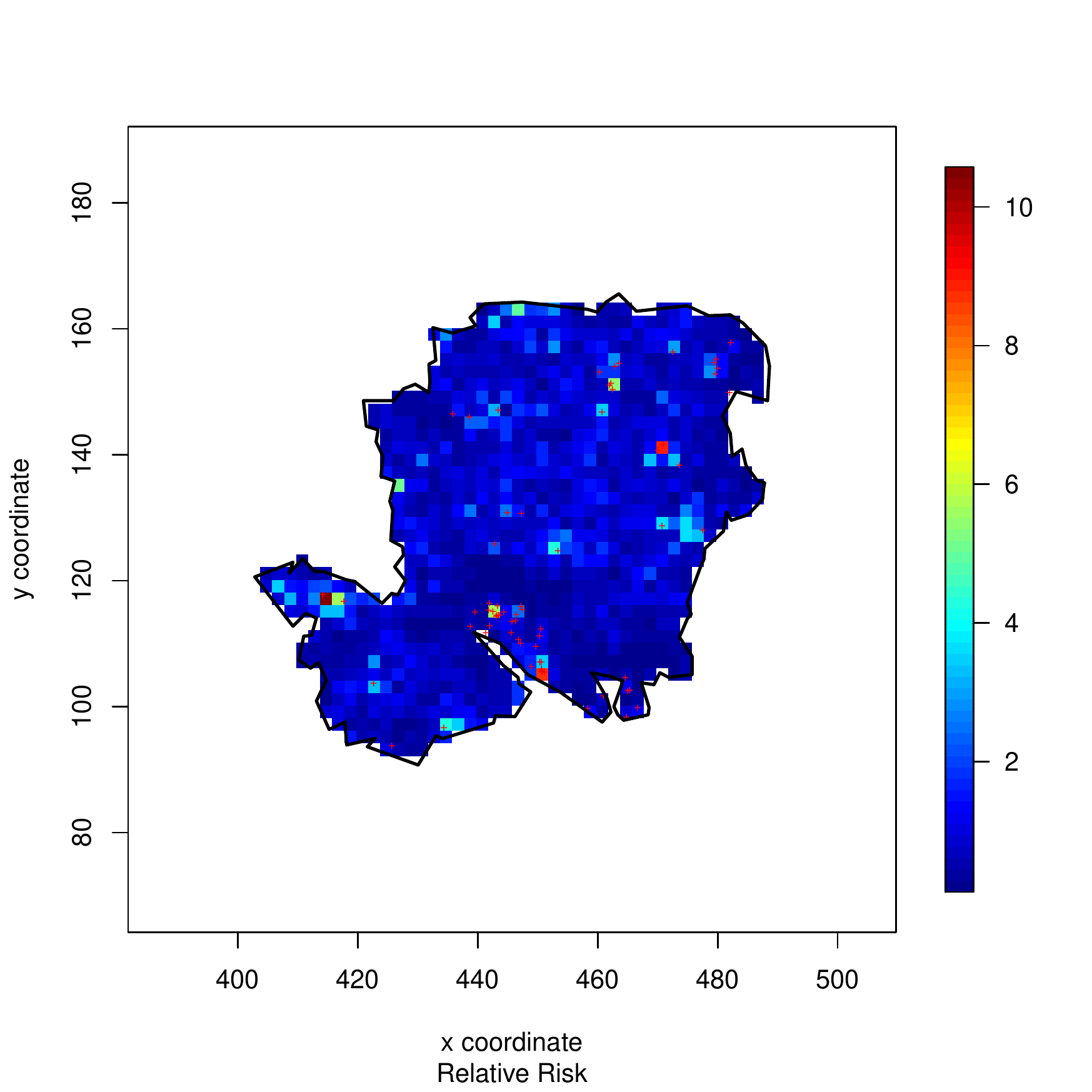}
   \end{minipage}\begin{minipage}{0.5\textwidth}
      \includegraphics[width=0.8\textwidth,height=0.8\textwidth]{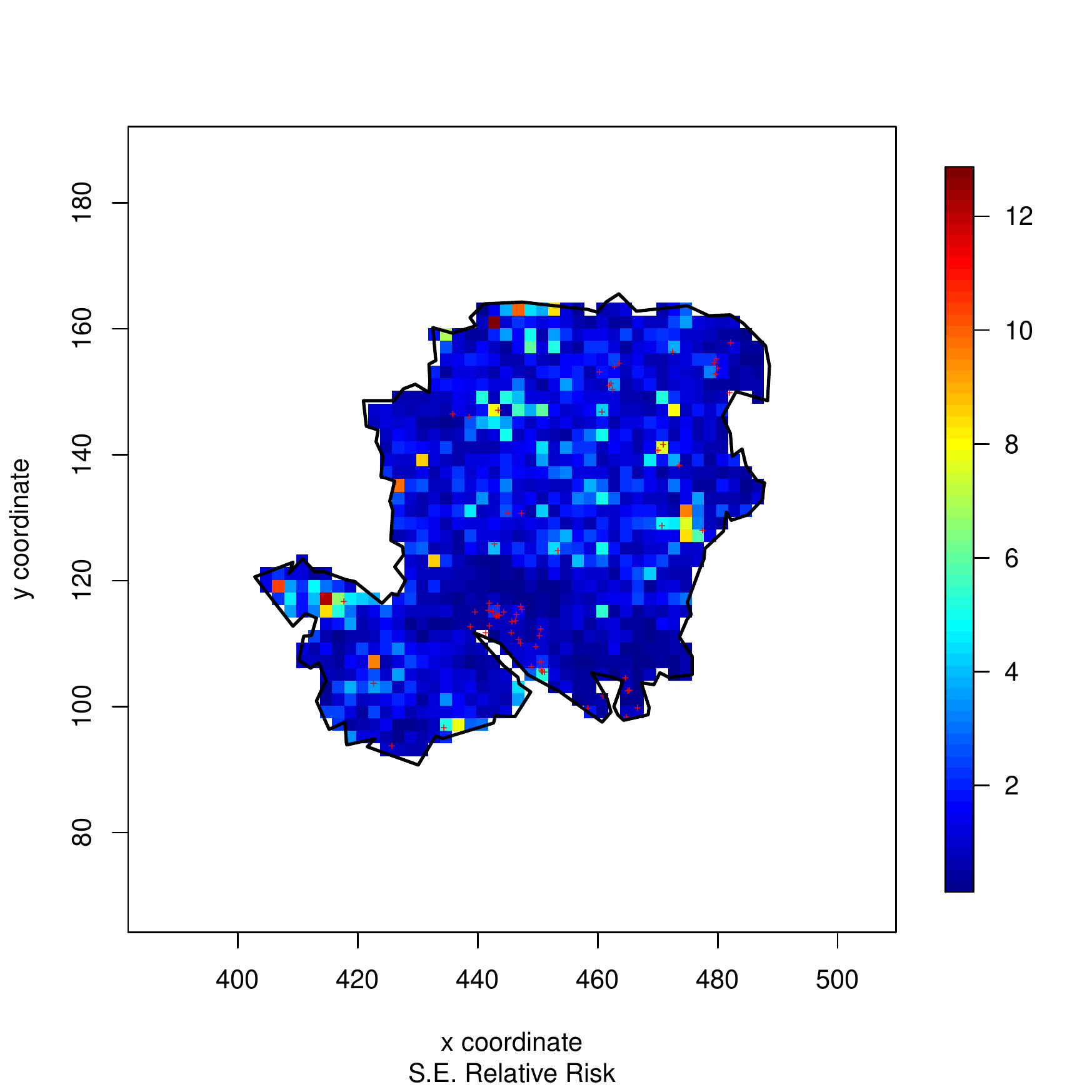}
   \end{minipage}
   \caption{\label{plotlg} Plots of the Monte Carlo mean relative risk (left) and associated standard errors (right).}
\end{figure}

Plots of the Monte Carlo mean relative risk and standard errors can be obtained with the commands:
\begin{CodeChunk}
\begin{CodeInput}
R> plot(lg,xlab="x coordinate",ylab="y coordinate")
R> plot(lg,type="serr",xlab="x coordinate",ylab="y coordinate")
\end{CodeInput}
\end{CodeChunk}
These commands produce a series of plots corresponding to each time step under
consideration; the plots shown in Figure~\ref{plotlg} are from the last time step,
time 50. 

To plot the mean Poisson intensity instead of the relative risk, the optional argument \code{type} can be set in the above: 
\begin{CodeChunk}
\begin{CodeInput}
R> plot(lg,type="intensity",xlab="x coordinate",ylab="y coordinate")
\end{CodeInput}
\end{CodeChunk}
The cases for each time step are also plotted by default.

\subsubsection{MCMC diagnostics}

\begin{figure}[htbp]
   \centering
   \begin{minipage}{0.05\textwidth}
   \hfill
   \end{minipage}\begin{minipage}{0.5\textwidth}
      \includegraphics[width=0.8\textwidth,height=0.8\textwidth]{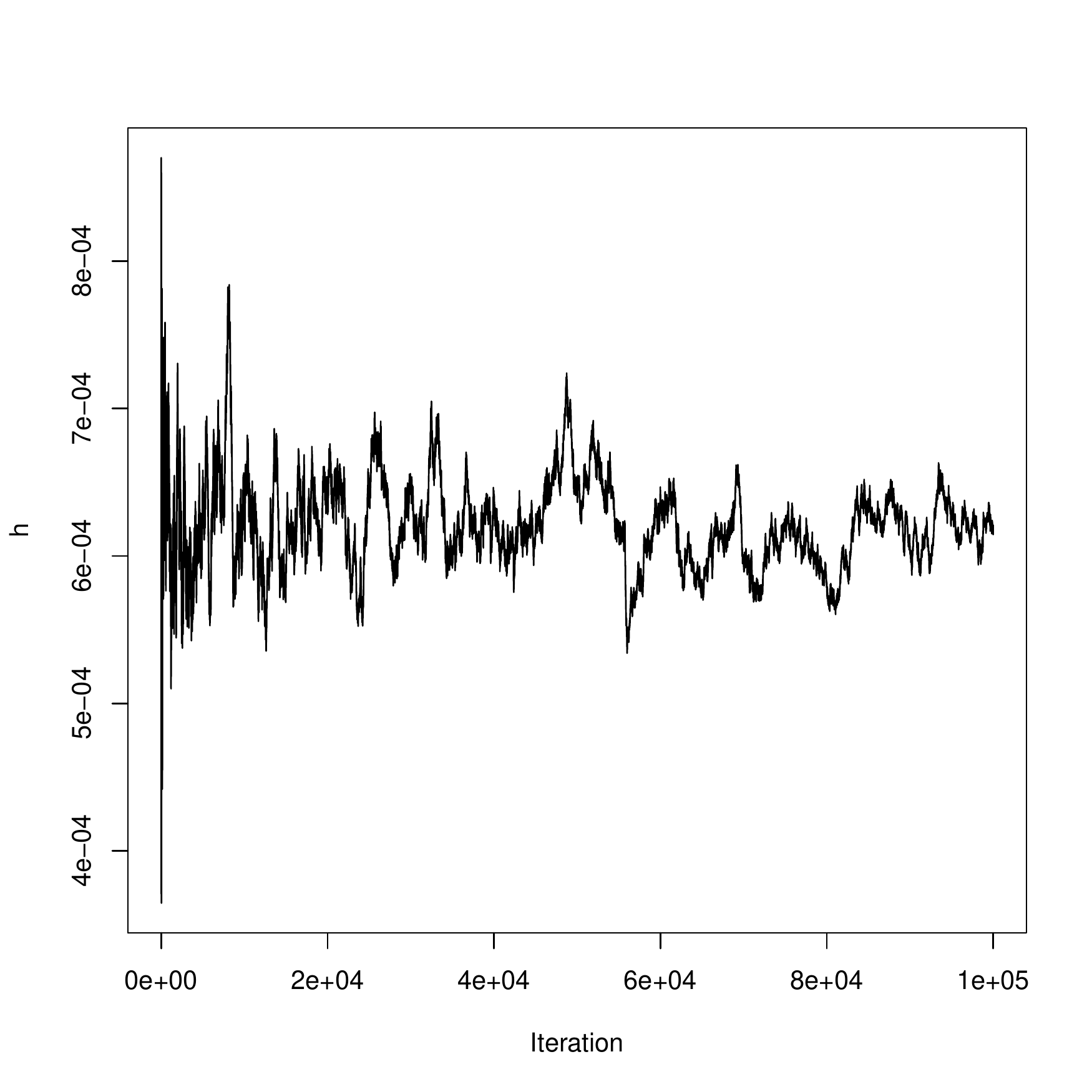}
   \end{minipage}\begin{minipage}{0.5\textwidth}
      \includegraphics[width=0.8\textwidth,height=0.8\textwidth]{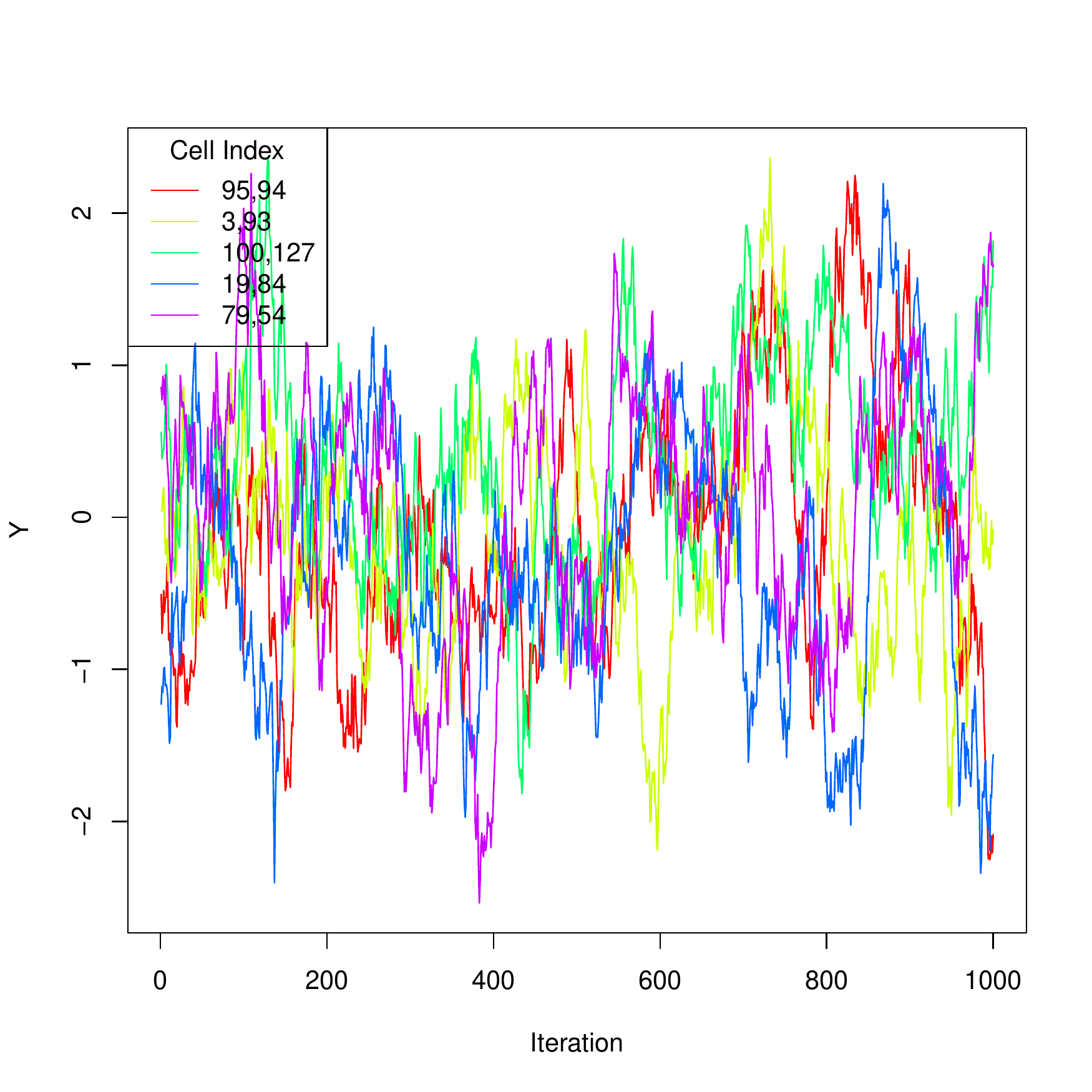}
   \end{minipage}
   \caption{\label{mcmcdiagnostics}MCMC diagnostic plots. Left: plot of values
of $h$ taken by the adaptive algorithm. Right: trace plots of the saved chains
from five grid cells.}
\end{figure}

MCMC diagnostics for the chain can either be based on a small sample of cells,
specified by the \code{MCMCdiag} argument of the \code{mcmcpars} function, or
via the full output from data dumped to disk (see
Section \ref{sect:netcdf}). The \code{hvals} command returns the value of $h$
used at each iteration in the algorithm, the left hand plot in Figure~\ref{mcmcdiagnostics} shows the values of $h$ for the non-burn-in period of the
chain; the adaptive algorithm was intialised with $h=1$,
which very quickly converged to around $h=0.0006$.

\begin{CodeChunk}
\begin{CodeInput}
R> plot(hvals(lg)[20000:120000],type="l",xlab="Iteration",ylab="h")
R> tr <- mcmctrace(lg)
R> plot(tr,idx=1:5)
\end{CodeInput}
\end{CodeChunk}

Trace plots, shown 
here in the right-hand panel of Figure~\ref{mcmcdiagnostics},
are also available using \code{plot(tr)} as in the above
code. Note that \code{mcmctrace} returns the trace for $\Gamma$. 
To plot the autocorrelation function, the standard \proglang{R} function can be
used.
For example,
\code{acf(tr$trace[,1])} gives the acf of the first saved chain.

\subsubsection{NetCDF}
\label{sect:netcdf}

The \pkg{lgcp} package
provides functions for accessing and performing computations on MCMC runs dumped
to disk. Because this can generate very large files, 
\pkg{lgcp} uses the cross-platform \code{NetCDF} file format for 
storage and rapid data access, as
provided by the package \pkg{ncdf} \citep{pierce2011}.
Access to subsets of these stored
data is via a file indexing system, which removes the need to load the complete
data into memory.

Subsets of data dumped to disk can be accessed with the \code{extract} function:
\begin{CodeChunk}
\begin{CodeInput}
R> subsamp <- extract(lg,x=c(4,10),y=c(32,35),t=6,s=-1)
\end{CodeInput}
\end{CodeChunk}
which returns an array of dimension $7\times4\times1\times1000$ (recall there were 1000 retained iterations). The arguments
\code{x} and \code{y} refer to the range of $x$ and $y$ \emph{indices} of the
grid of interest whilst \code{t} 
specifies the time points of interest. Note, however,
that in this example times 45 through 50 were used for prediction, 
and \code{t=6} here refers to the sixth of these time-points,
i.e., time 50. Finally,
\code{s=-1} stipulates that all simulations are to be returned. 
More generally, each argument of
\code{extract}
can be specified either
as a range or set equal to $-1$, in which case all of the data in that
dimension are returned. 
The \code{extract}
   function can also extract MCMC traces from individual cells using,
   for example,
   \code{extract(lg,x=37,y=12,t=6)}.

Should the user wish to extract data from a polygonal subregion of the
observation window, this can be achieved with the command
\begin{CodeChunk}
\begin{CodeInput}
R> subsamp2 <- extract(lg,inWindow=win2,t=6)
\end{CodeInput}
\end{CodeChunk}
where \code{win2} is a polygonal observation window defined below. 
Here, \code{win2} had been selected using the following commands:
\begin{CodeChunk}
\begin{CodeInput}
R> plot(window(lg))
R> win2 <- loc2poly()
\end{CodeInput}
\end{CodeChunk}
The first of the above commands plots the observation window,
whilst
the second is a wrapper function for the \proglang{R} function \code{locator}.
When invoked, \code{loc2poly()} allows the user to select areas of the
observation window manually from the graphics device opened by the first
command: the user simply makes a series of left clicks, traversing the required
window in a single direction (i.e., clockwise \emph{or} anticlockwise), finishing
the polygon with a right click.
The resulting selection is converted into a \pkg{spatstat} polygonal \code{owin}
object. The user could also 
specify the \code{extract} argument \code{inWindow} directly using a \pkg{spatstat} \code{owin} object.

If the user decides that some other summary 
than those specified by the 
\code{gridmeans} option is of interest, 
this can easily be computed
from the stored data (c.f., Section \ref{sect:outputspec})
The syntax is then slightly different,
as in the following example that
computes the same exceedances in Section \ref{sect:outputspec}:

\begin{CodeChunk}
\begin{CodeInput}
R> ex <- expectation(obj=lg,fun=exceed)
\end{CodeInput}
\end{CodeChunk}

Alternatively, cell-wise quantiles of \emph{functions} 
of the stored data can also be retrieved and plotted:
\begin{CodeChunk}
\begin{CodeInput}
R> qt <- quantile(lg,c(0.5,0.75,0.9),fun=exp)
R> plot(qt,xlab="X coords",ylab="y coords")
\end{CodeInput}
\end{CodeChunk}
As for the extract function above, quantiles can also 
be computed for smaller spatial observation windows. The indices of any cells of
interest in these plots can be retrieved by typing \code{identify(lg)}.
Cells are then selected via left mouse clicks in the graphics device, selection
being terminated by a right click.

\begin{figure}[htbp]
   \centering
   \begin{minipage}{0.05\textwidth}\hfill    
   \end{minipage}\begin{minipage}{0.5\textwidth}
      \includegraphics[width=0.8\textwidth,height=0.8\textwidth]{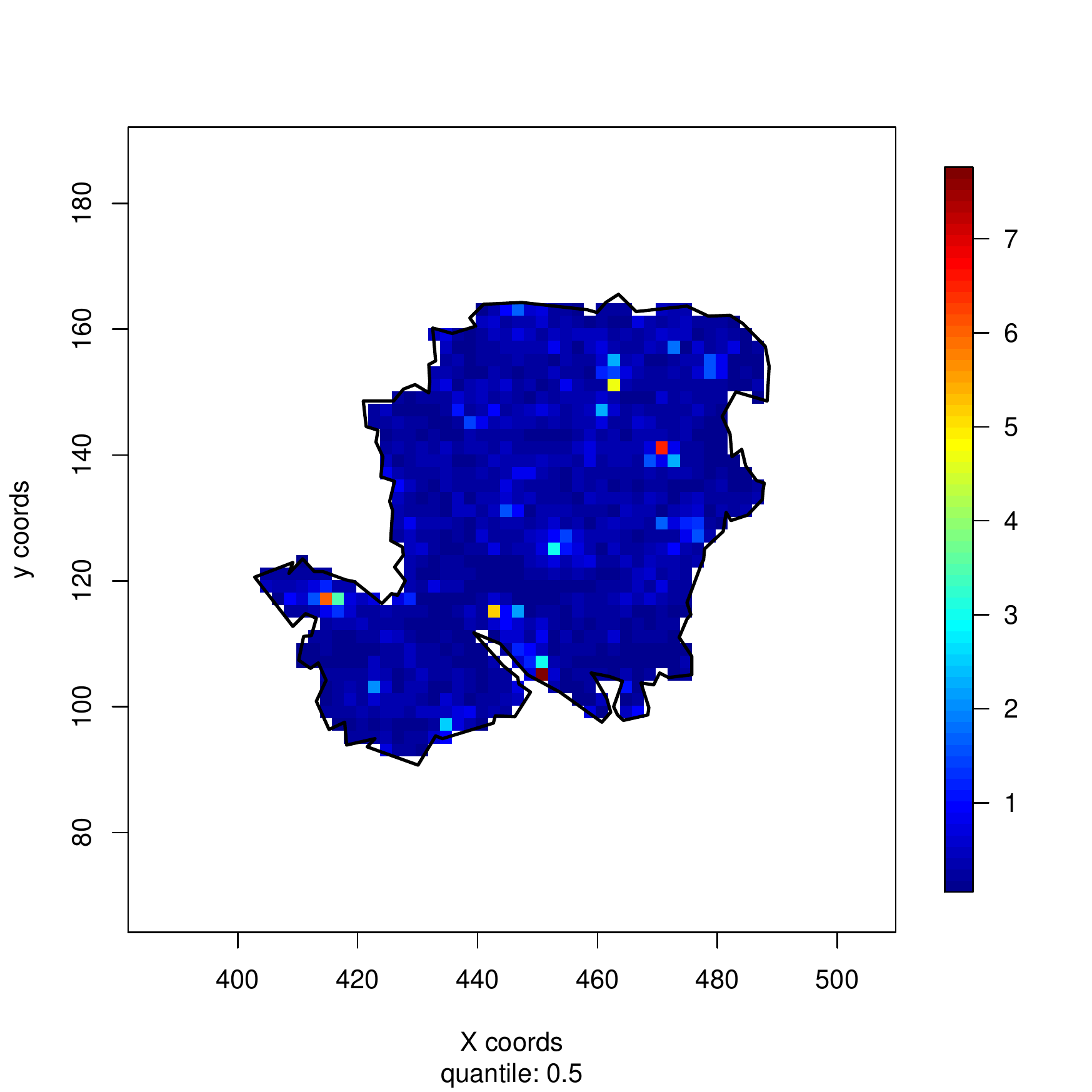}
   \end{minipage}\begin{minipage}{0.5\textwidth}
      \includegraphics[width=0.8\textwidth,height=0.8\textwidth]{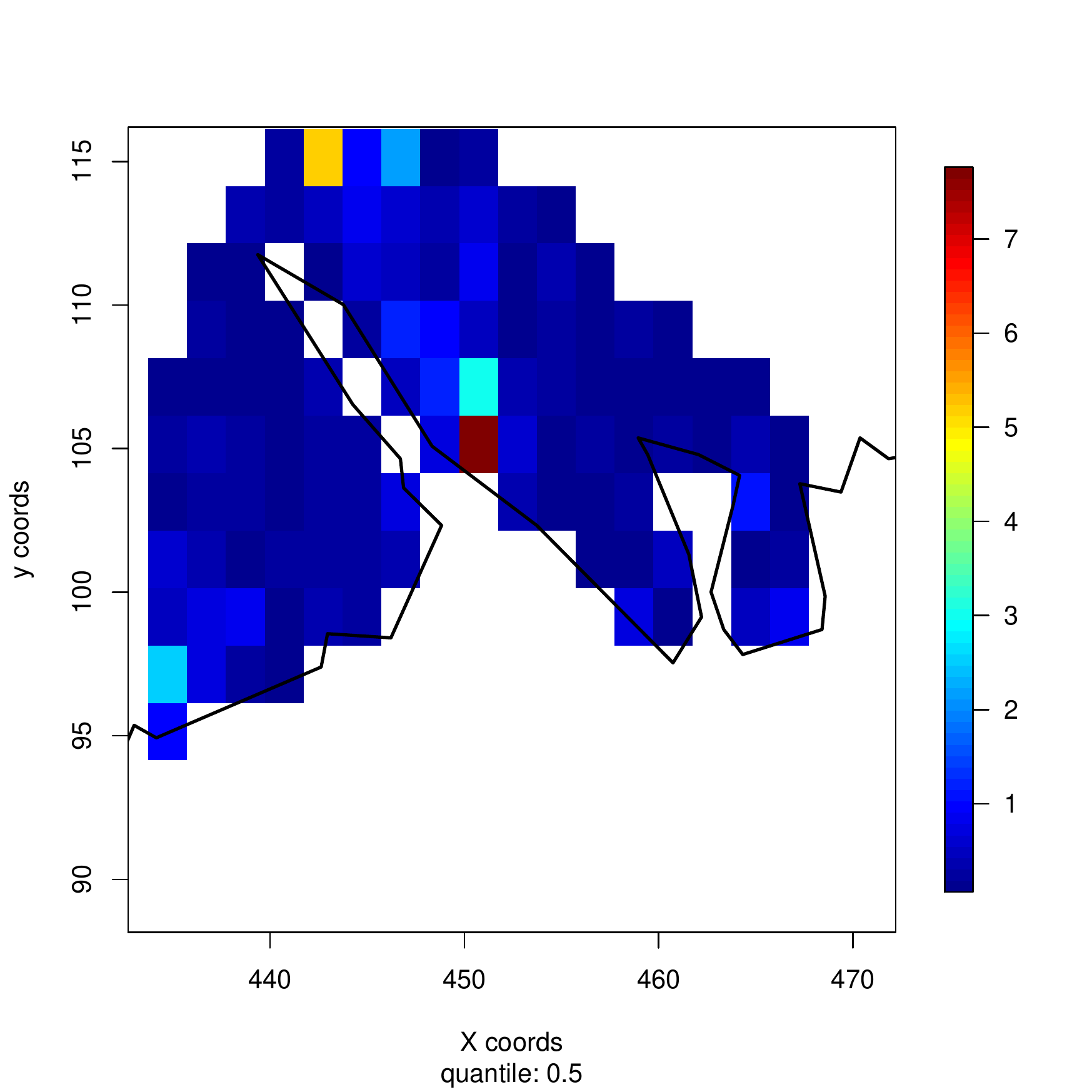}
   \end{minipage}
   \caption{\label{qtile} Plot showing the median 
   of relative risk (obtained using \code{fun=exp} as in the text) computed from
the simulation. Left: quantiles computed for whole window. Right: 
   zooming in on the lower area of the map, representing the
   cities of Southampton and Portsmouth. Greater detail is available by
initially performing the simulation on a finer grid.}
\end{figure}

\begin{figure}[htbp]
   \centering

\includegraphics[width=0.286\textwidth,height=0.225\textwidth]{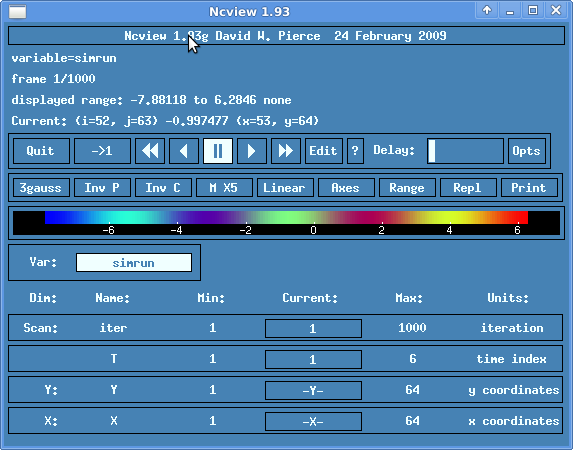}

\includegraphics[width=0.164\textwidth,height=0.173\textwidth]{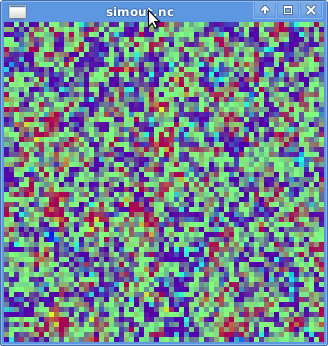}

\includegraphics[width=0.334\textwidth,height=0.196\textwidth]{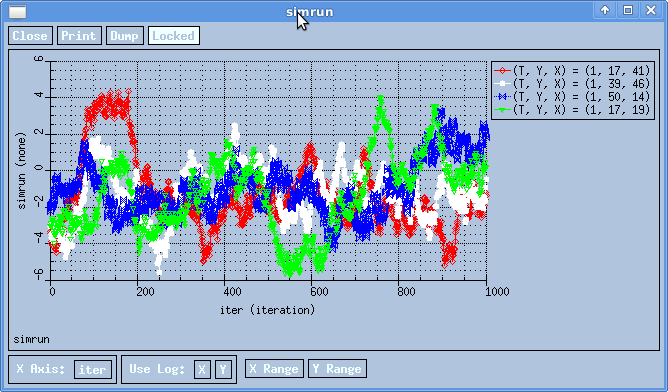}

\caption{\label{netview} Viewing a MALA run with software \code{netview}.}
\end{figure}

Lastly, Linux users can benefit from the software \proglang{Ncview}, available
from \url{http://meteora.ucsd.edu/~pierce/ncview_home_page.html},
which provides fast visualisation of NetCDF files. 
Figure~\ref{netview} shows a screen-shot, with the control panel (left), an
image of one of the sampled grids (middle) and several MCMC chains (right),
which are obtained by clicking on the sampled grids; up to five
chains can be displayed at a time. 
There are equivalent tools for Windows users.

\subsubsection{Plotting exceedance probabilities}

\begin{figure}[htbp]
   \centering
   \includegraphics[width=0.5\textwidth,height=0.5\textwidth]{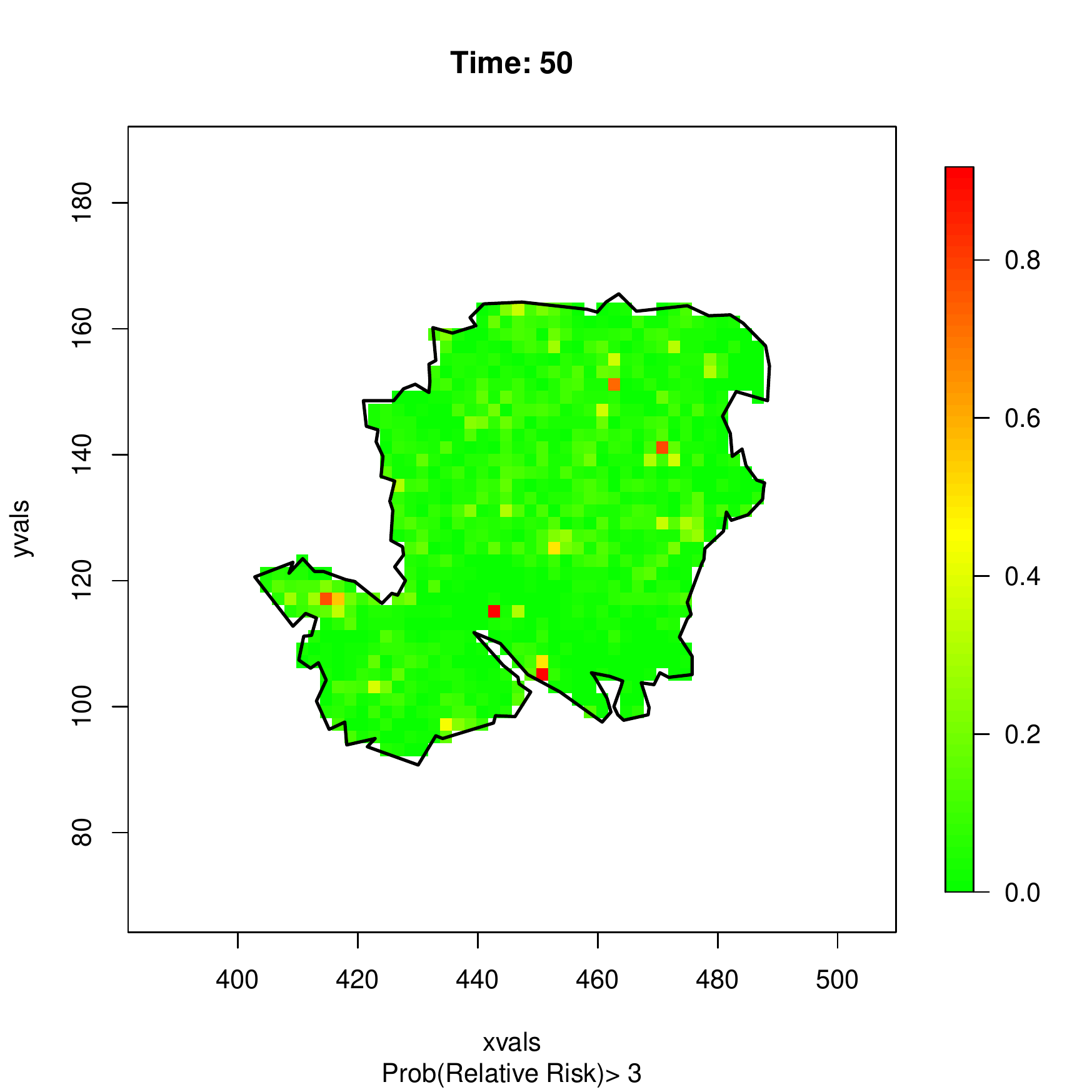}
   \caption{\label{exceedance} Plot showing the cellwise probability (colour coded) that the relative risk is greater than 3.}
\end{figure}

Recall that
the object \code{exceed}, defined above,
was a function with an attribute giving a vector of thresholds to compute
cell-wise exceedance probabilities at each threshold. 
A plot can be produced either directly from the \code{lgcpPredict} 
object,
\begin{CodeChunk}
\begin{CodeInput}
R> plotExceed(lg, fun = "exceed")
\end{CodeInput}
\end{CodeChunk}
or, equivalently, from 
the output of an \code{expectation} on an object dumped to disk:
\begin{CodeChunk}
\begin{CodeInput}
R> plotExceed(ex[[6]], fun = "exceed",lgcppredict=lg)
\end{CodeInput}
\end{CodeChunk}
Recall also
that the option \code{lastonly=TRUE} was selected for \code{MonteCarloAverage}, 
hence \code{ex[[6]]} in the second example above 
corresponds to the same set of plots as for the first example. The advantage of
computing
expectations from files dumped to 
disk is flexibility. For example,
if the user now wanted to plot the exceedances for day 49, 
this is simply achieved by replacing \code{ex[[6]]} with \code{ex[[5]]}. 
Also, 
exceedances for a  new set of thresholds can be computed 
by creating,
for example, a new function by the command
\code{exceed2 <- exceedProbs(c(2.3,4))}. 
An example of the resulting
output is given in Figure~\ref{exceedance}.

\section{Future extensions}
\label{sect:extensions}

This article has decribed how \pkg{lgcp} may be used to fully specify, fit, and simulate from i.e., predict (both unconditionally and conditionally, dependent on observed data) a spatio-temporal log-Gaussian Cox process on $\mathbb{R}^2$. A substantial volume of novel code, such as access to fast-Fourier transform methods needed in simulation and the first open-source \proglang{R} implementation of the Metropolis-adjusted Langevin algorithm for this target, has been instrumental in the development of this package.

The initial motivation for this work was disease surveillance as performed by \cite{diggle2005}, and it is this application which has driven the core functionality for initial release (at the time of writing, \pkg{lgcp} is at version 0.9-0). However, there is much scope for potential extensions. Originally introduced into spatial statistics by \cite{moller1998}, the ability to analyse `purely spatial' data sets with the very flexible log-Gaussian Cox Process is seen as one of the most pressing additional developments soon to be included.

A list of other potential extensions to \pkg{lgcp} includes: the ability to use covariate information; to perform principled Bayesian parameter inference (currently under investigation); and to handle applications where covariate information is available at differing spatial resolutions. Finally, in the spatio-temporal setting, it is of further interest to include the ability to handle non-separable correlation structures; see for example \cite{gneiting2002,rodrigues2010}. 

\section{Acknowledgements}

The population data used in this article was based on real data from project AEGISS \citep{diggle2005}. AEGISS was supported by a grant from the Food Standards Agency, U.K., and from the National Health Service Executive Research and Knowledge Management Directorate.

\appendix

\section{Rotation}
\label{sect:rotation}

The MALA algorithm works on a regular square grid 
placed over the observation window. The user is
responsible for providing a physical grid size 
on which to perform estimation/prediction.
The gridded observation window is then extended automatically to obtain a $2^m\times2^n$ grid on which 
the simulation is performed. 
By default, the orientation of this extended grid is 
the same as the object \code{win}. If the observation window is elongated and
set at a diagonal, then some loss of efficiency 
that would occur as a consequence
of
redundant computation at irrelevant locations can be recovered 
by rotating
the coordinate axes and performing the computations 
in the rotated space.

\begin{figure}[htbp]
   \centering
   \begin{minipage}{0.05\textwidth}
   \hfill
   \end{minipage}\begin{minipage}{0.5\textwidth}
      \includegraphics[width=0.8\textwidth,height=0.8\textwidth]{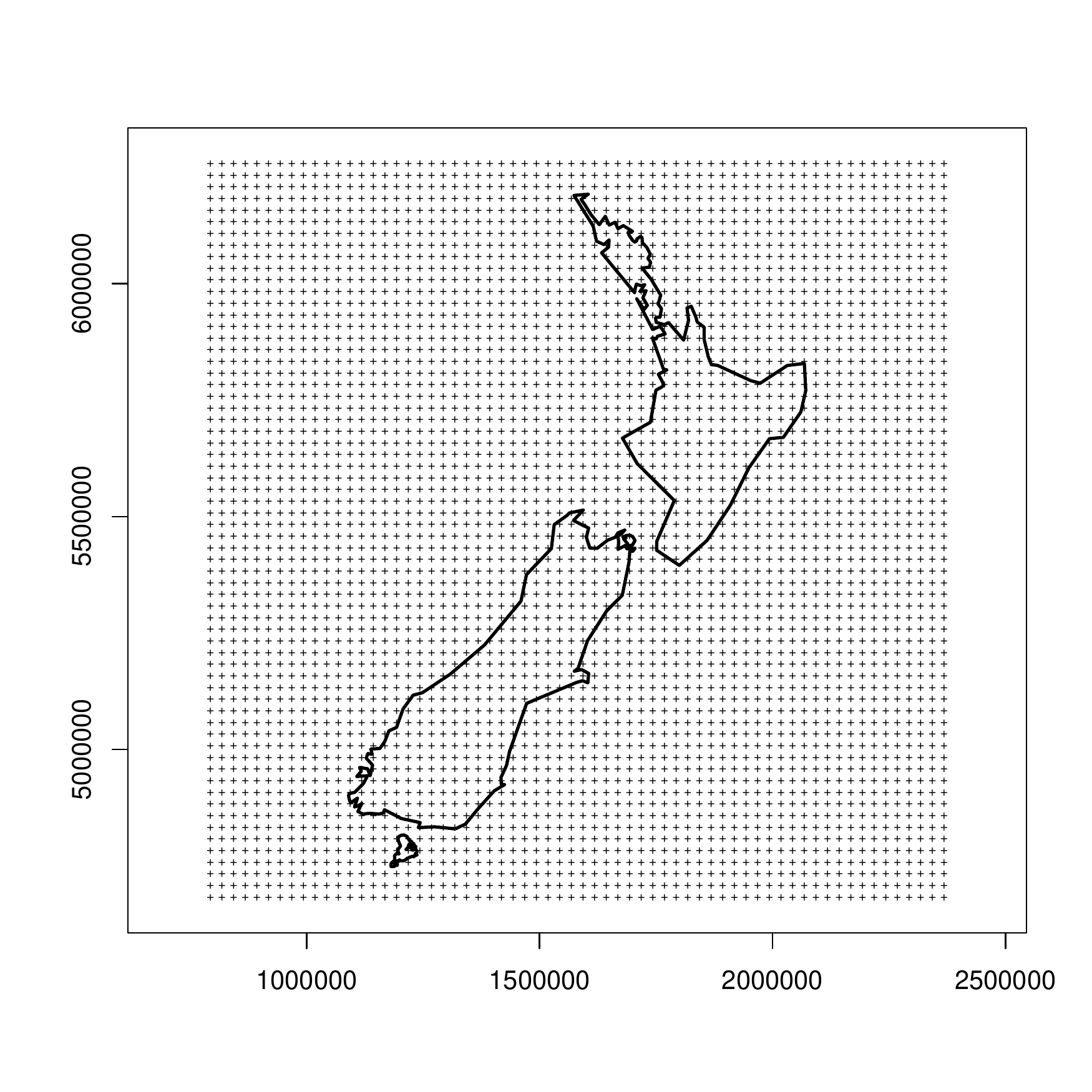}
   \end{minipage}\begin{minipage}{0.5\textwidth}
      \includegraphics[width=0.8\textwidth,height=0.8\textwidth]{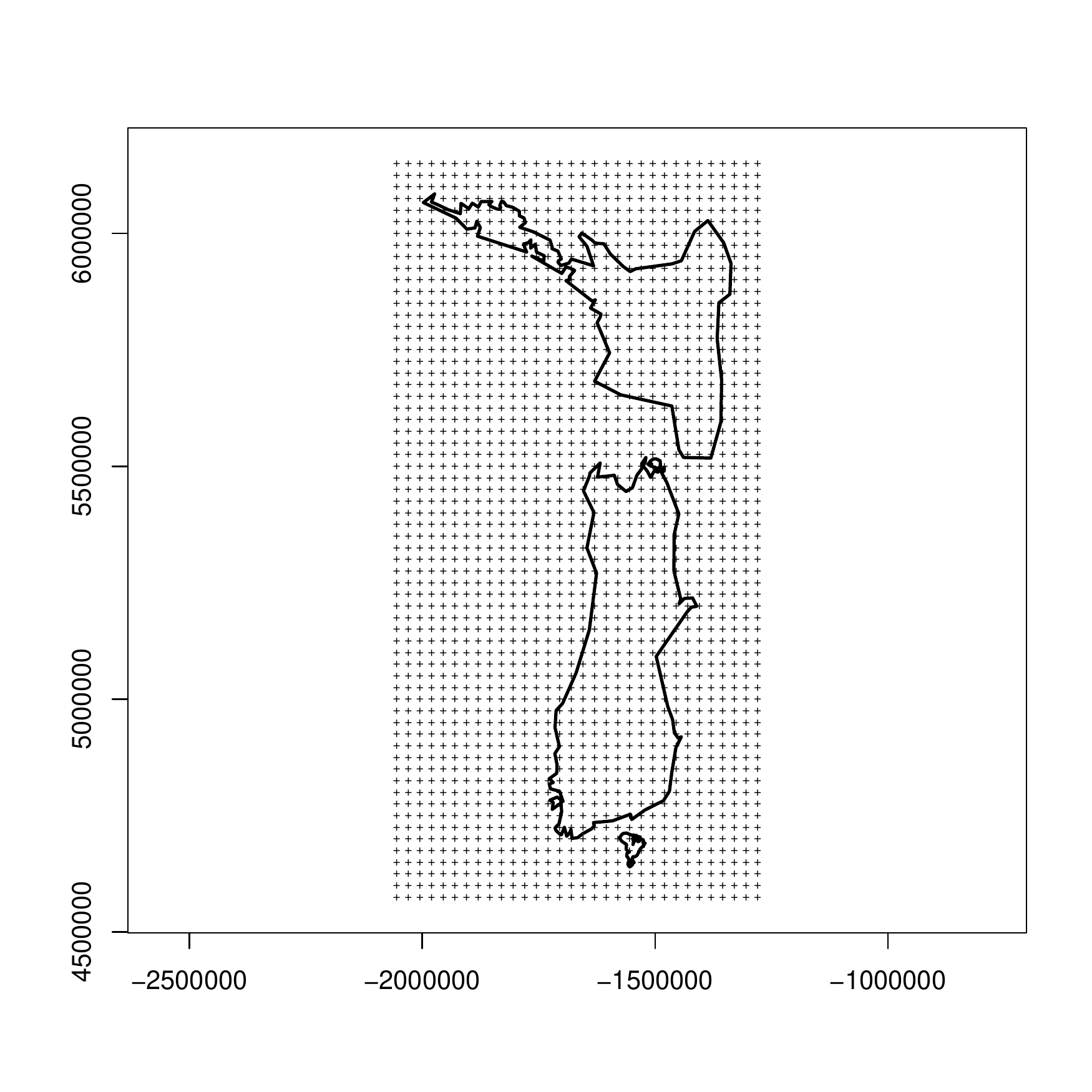}
   \end{minipage}
   \caption{\label{roteffgain} Illustrating the potential gain in efficiency by
rotating the observation window. Left plot: the selected grid without rotation.
Right plot: the optimally rotated grid.}
\end{figure}

To illustrate this, suppose \code{xyt2} is an \code{stppp} object with such an
elongated and diagonally oriented
window (see Figure~\ref{roteffgain}). The function \code{roteffgain} displays 
whether any efficiency can be gained by rotation; clearly this not only depends
on the observation window, but also on the size of the square cells on which the
analysis will be performed. In the example below, the user wishes to perform the
analysis using a cell width of 25km 
(corresponding to \code{cellwidth=25000} in the code below):
\begin{CodeChunk}
\begin{CodeInput}
R> roteffgain(xyt2,cellwidth=25000)
\end{CodeInput}
\begin{CodeOutput}
By rotating the observation window, the efficiency gain would be: 200%, 
   see ?getRotation.stppp
NOTE: efficiency gain is measured as the percentage increase in FFT 
   grid cells from not rotating compared with rotating
[1] TRUE
\end{CodeOutput}
\end{CodeChunk}
The routine returns
\code{FALSE} if there is no `efficiency gain'. Note that the efficiency gain is
not a reflection on computational speed, but rather a measure of how many fewer
cells the MALA is required to use; this is illustrated in Figure~\ref{roteffgain}. As a technical aside, a better measure would be a ratio of
mixing times for the MCMC chains based on unrotated and rotated windows;
however, as the mixing time depends on how well the MALA has been tuned, it is
not clear how this can be estimated accurately.

Having ascertained whether rotation is advantageous, the optimally rotated data,
observation window and rotation matrix 
can be retrieved using the function \code{getRotation}. For
prediction
using \code{lgcpPredict}, there is also an \code{autorotate} option: this
allows the user to perform MALA on a rotated grid with minimal input so long as
rotation leads to a gain in efficiency. If the model is fitted using a rotated
frame, then the predictions will also be returned in this frame: this means that
in the original orientation the output will be on a grid misaligned to the original axes. 
The \pkg{lgcp} package
provides methods for the generic function \code{affine} so that \code{stppp}
and \code{spatialAtRisk} objects can be rotated manually.

\section{Simulating data}
\label{sect:simulation}

The \pkg{lgcp} package
also provides an approximate simulation tool for drawing samples from the model in Equation~\ref{eqn:themodel}. Simulation minimally requires an observation window, a
range of times over which to simulate data, spatial and temporal intensity
functions $\lambda$ and $\mu$, a cell width for the discretisation  and a set of
spatial/temporal model parameters together with a choice of spatial covariance
model.

The code below simulates data from a log-Gaussian Cox process on the observation
window from the example in the text above. The function \code{tempfun} is coerced into a \code{temporalAtRisk} object and defines
the temporal trend. 
Any appropriately defined \code{temporalAtRisk} object can be used here.
Similarly, \code{spatial.intensity} can either be an object of class
\code{spatialAtRisk} or one that can be coerced to one.
\begin{CodeChunk}
\begin{CodeInput}
R> W <- xyt$window
R> tempfun <- function(t){return(100)}
R> sim <- lgcpSim(owin=W,
	       tlim=c(0,100), 
	       spatial.intensity=den,
	       temporal.intensity=tempfun,
	       cellwidth = 0.5,
	       model.parameters=lgcppars(sigma=2,phi=5,theta=2))
\end{CodeInput}
\end{CodeChunk}
Note that
the finer the grid resolution, the more accurately will
the process  be simulated, and that smaller values of $\phi$
require a finer
discretisation. A warning is issued if the algorithm 
thinks the chosen cell width is too large. 
The disctretisation in time is chosen automatically by the algorithm. 

\section[Handling the SpatialAtRisk class]{Handling the \code{SpatialAtRisk}
class}
\label{sect:spatialAtRisk}

\begin{figure}[htbp]
   \centering
   \includegraphics[width=0.8\textwidth,height=0.6\textwidth]{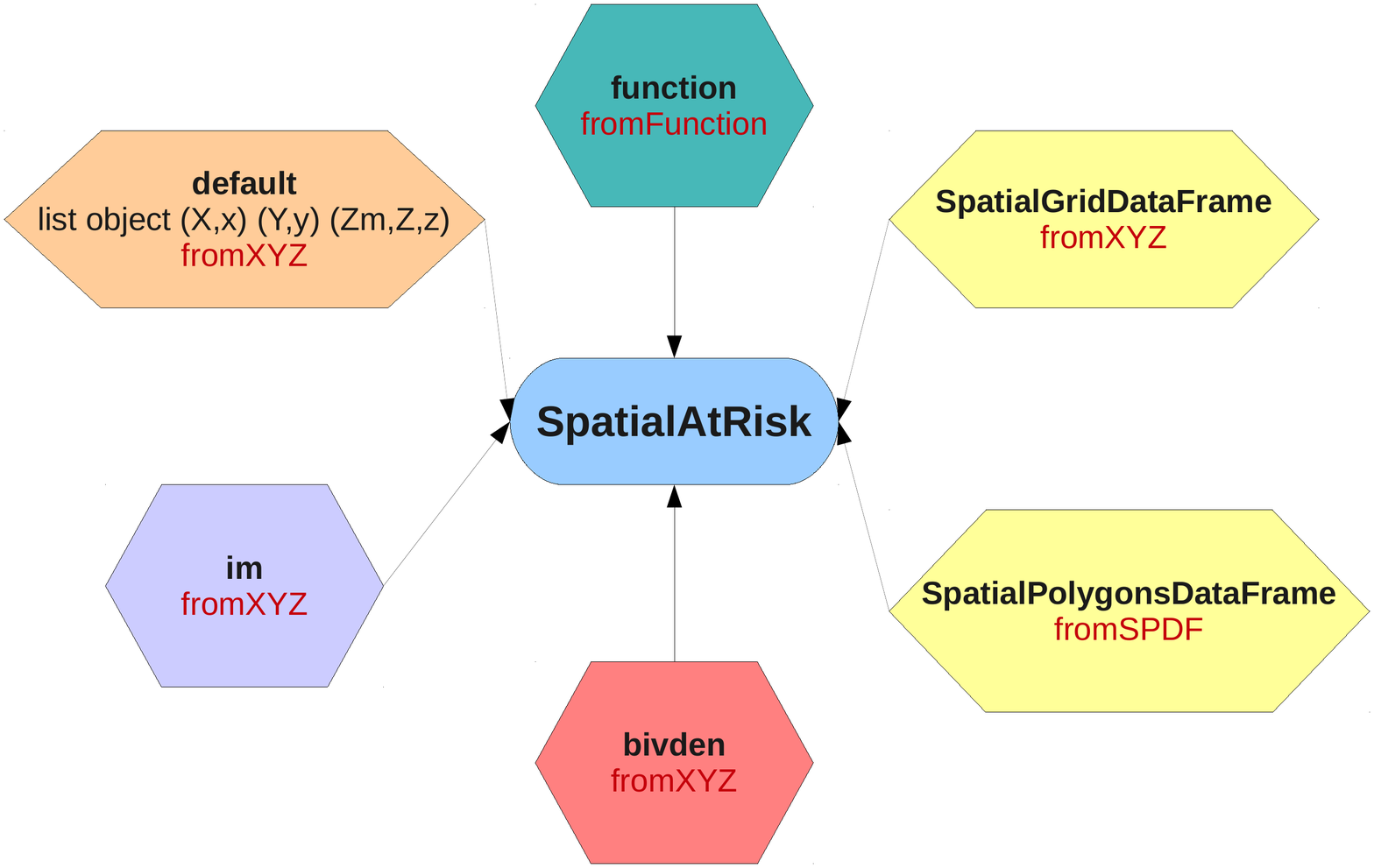}
   \caption{\label{conversion1}Conversion to \code{spatialAtRisk} objects. By
default, \code{SpatialAtRisk} looks for a \code{list}-type object, but other
objects that can be coerced include \pkg{spatstat} \code{im} objects,
\code{function} objects, \pkg{sp} \code{SpatialGridDataFrame} and
\code{SpatialPolygonsDataFrame} objects and \pkg{sparr} \code{bivden} objects.
The text in red gives the type of \code{spatialAtRisk} object created.}
\end{figure}

\begin{figure}[htbp]
   \centering
   \includegraphics[width=0.6\textwidth,height=0.4\textwidth]{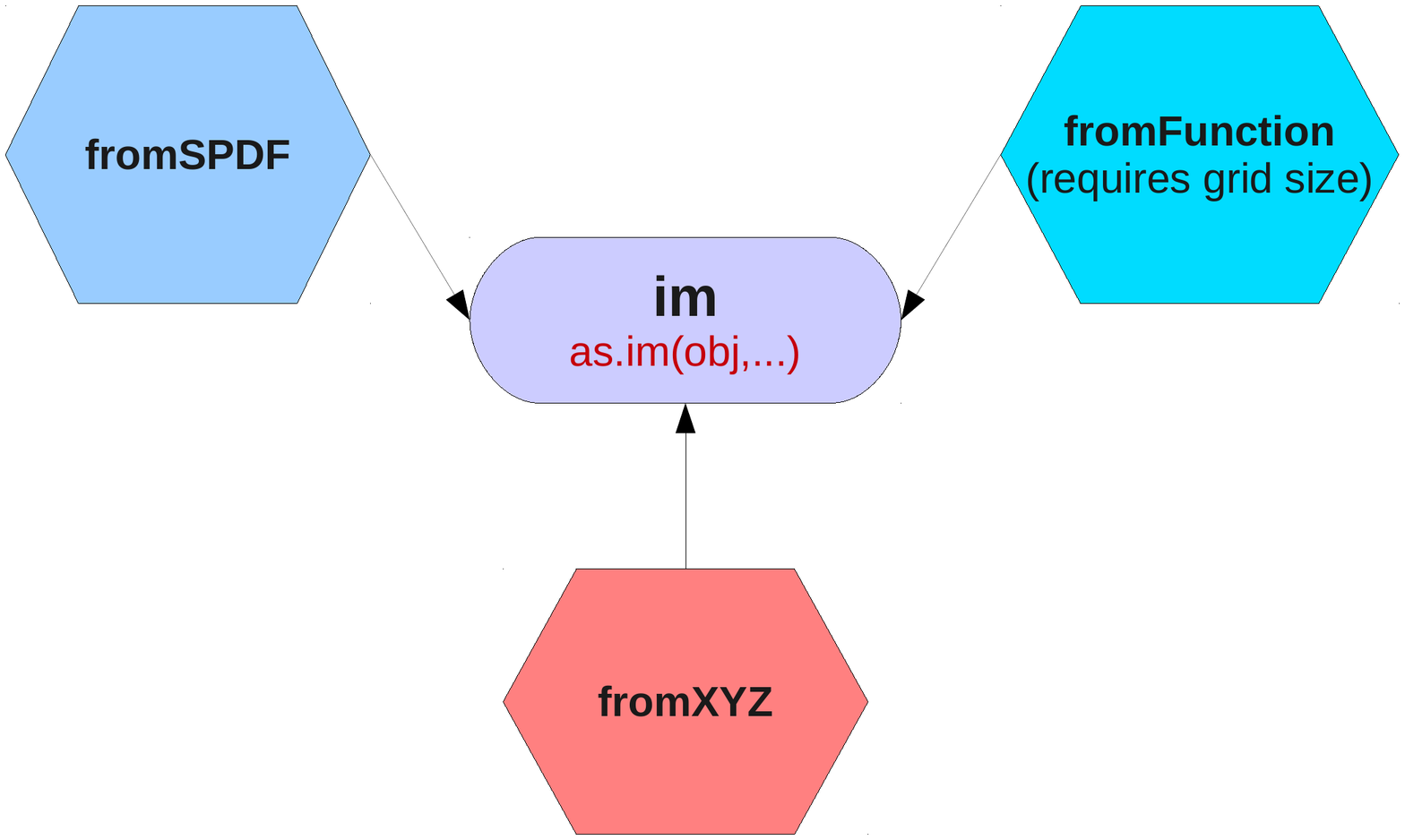}
   \caption{\label{conversion2}Conversion to \pkg{spatstat} objects of class
\code{im}; these are useful for parameter estimation, in each case a call to the
function \code{as.im(obj,...)} will perform the coercion.}
\end{figure}

This section illustrates the available commands for converting between different
types of \proglang{R} objects that can be used to describe $\lambda(s)$.
Conversion methods are provided for objects from the packages \pkg{spatstat}
\citep{baddeley2005}, \pkg{sp} \citep{bivand2008} and \pkg{sparr}
\citep{davies2011}.
These are illustrated in Figure~\ref{conversion1}. For the purposes of parameter
estimation, Figure~\ref{conversion2} shows the different \code{spatialAtRisk}
objects that can be converted into 
an appropriate format (i.e., a \pkg{spatstat} \code{im} object).

There is also a function to convert from \code{fromXYZ}-type
\code{spatialAtRisk} objects to \pkg{sp} objects of class \code{SpatialGridDataFrame}:
\code{as.SpatialGridDataFrame(obj,...)}. Lastly, \code{fromFunction}-type can be
converted to \code{fromXYZ}-type \code{spatialAtRisk} objects using the
\code{as.fromXYZ} function. Note that if a \code{spatialAtRisk} object is
specified via a function, then it is the user's responsibility to ensure that
the function integrates to 1 over the observation window; one way to bypass this
problem is to convert the function to an \code{spatialAtRisk} object of
\code{fromXYZ}-type.

\newpage
\bibliographystyle{jss}
%\bibliography{../bibliography/bibliography}
\bibliography{LGCP_bibliography}

\end{document}